\documentclass[times]{MyArticle}

\usepackage[colorlinks,bookmarksopen,bookmarksnumbered,citecolor=blue,urlcolor=blue]{hyperref}

\usepackage{mathrsfs}
\usepackage[usenames,dvipsnames,svgnames,table]{xcolor}

\usepackage{mathptmx}      % use Times fonts if available on your TeX system
\usepackage{subfig}
\DeclareMathAlphabet{\mathcal}{OMS}{cmsy}{m}{n}
\SetMathAlphabet{\mathcal}{bold}{OMS}{cmsy}{b}{n}
\begin{document}

\title{A Meshfree Generalized Finite Difference Method for Solution Mining Processes}

%\author{Pratik Suchde \affil{1}\corrauth, ABC \affil{1}, DEF \affil{2}}

%\address{\affilnum{1}Fraunhofer ITWM, Fraunhofer-Platz 1, 67663 Kaiserslautern, Germany. \break %
%\affilnum{2}So and so place.}

\corraddr{E-mail: isabel.michel@itwm.fraunhofer.de}

\author{Isabel Michel  \affil{1}\corrauth,  Tobias Seifarth \affil{1},  J\"org Kuhnert  \affil{1},  Pratik Suchde \affil{1}}

\address{Fraunhofer Institute for Industrial Mathematics ITWM, 
           Fraunhofer-Platz 1, 67663 Kaiserslautern, Germany
          }

%\date{Received: date / Accepted: date}
% The correct dates will be entered by the editor

\begin{abstract}
Experimental and field investigations for solution mining processes have improved intensely in recent years. Due to today's computing capacities, three-dimensional simulations of potential salt solution caverns can further enhance the understanding of these processes. They serve as a ``virtual prototype" of a projected site and support planning in reasonable time. In this contribution, we present a meshfree Generalized Finite Difference Method (GFDM) based on a cloud of numerical points that is able to simulate solution mining processes on microscopic as well as macroscopic scales, which differ significantly in both the spatial and temporal scale. Focusing on anticipated industrial requirements, Lagrangian and Eulerian formulations including an Arbitrary Lagrangian-Eulerian (ALE) approach are considered.
\keywords{Meshfree Methods ; Generalized Finite Difference Method ; Lagrangian Formulation ; Arbitrary Lagrangian-Eulerian Formulation ; Solution Mining}
\end{abstract}

\maketitle

\vspace{-6pt}

%%%%%%%%%%%%%%%%%%%%%%%%%%%%%%%%%%%%%%%%%%%%%%%%%%
%%%%%%%%%%%%%%%%%%%%%%%%%%%%%%%%%%%%%%%%%%%%%%%%%%
%%%
\section{Introduction}

\begin{figure}
  \centering
  \includegraphics[width=0.55\textwidth]{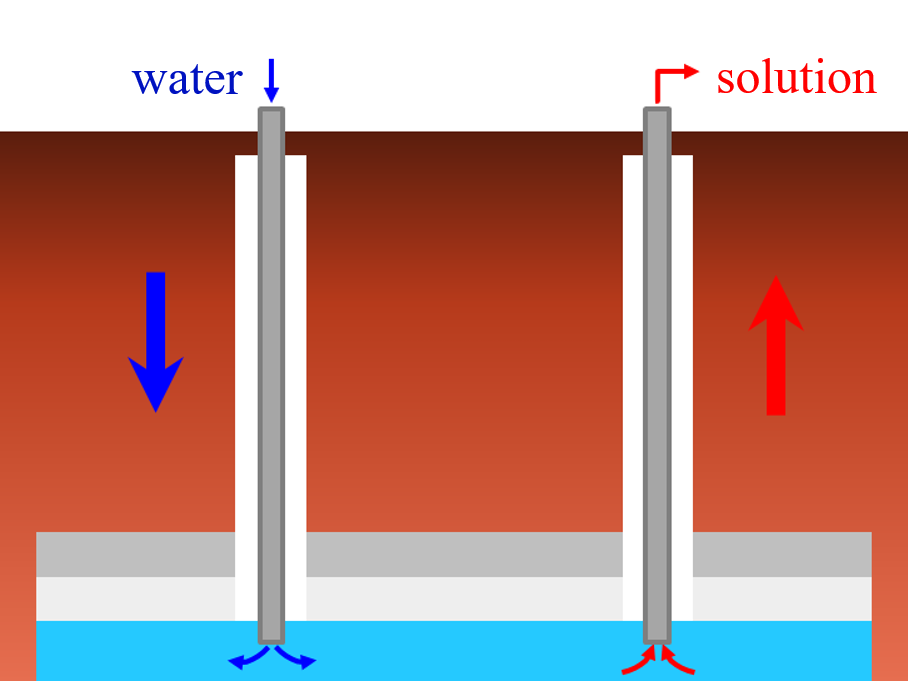}
\caption{Schematic of double-well solution mining (adapted from \cite{Seifarth2018}).}
\label{fig:SchematicDoubleWellSolutionMining}
\end{figure}

The basic motivation of this research is to provide a method that is able to simulate the long-term development of a salt cavern during a double-well solution mining process. Solution mining is used to extract underground water-soluble minerals such as salt and potash. A double-well convection process has been a preferred choice for solution mining due to it's large recovery rate \cite{Chen2020,Zhang2018}. As the name suggests, this involves the use of two boreholes or wells for the extraction process: an injection well, and a recovery or extraction well. For the extraction of salt, fresh water is pumped into a salt deposit through the first ``injection" well. Salt present in the cavern dissolves in the water to produce a saturated brine solution. This is then extracted at the second ``extraction" well. A schematic of this process is shown in Fig.~\ref{fig:SchematicDoubleWellSolutionMining}. The main direction of dissolution is vertical, which is controlled by alternate lifting of the injection and the extraction well. 

In this work, we focus on modeling of the fluid flow involved in such a double-well solution mining procedure, including the formation of the salt-water solution. An essential aspect of this is to accurately model the long-term geometrical evolution of the salt cavern. This is needed to steer the actual process of solution mining, in terms of, for example, determining when and at what rate the injection and extraction wells are raised. However, numerically modeling this is very challenging, as it is a highly dynamic three-dimensional process involving different spatial and temporal scales. Over the time scale of several years, as salt in the cavern dissolves in the water, the cavern starts to erode, causing significant deformations in it's overall shape. However, the dissolution process relies on a smaller time scale of several minutes. On the spatial scale, the former involves the modeling of the entire salt cavern, while the latter is more localized and is relevant near the cavern walls. In the present work, we model both these processes in separate simulation setups. A \emph{macroscopic} simulation is carried out to model the evolution of the cavern over many years. This is done on actual salt cavern geometries. The computation of the diffusion rate of the salt (and related minerals) to be used in these macroscopic simulations are done in a separate simulation, in the so-called \emph{microscopic} setup. This involves simulations over the smaller time scale of a few minutes, and over representative geometries several orders of magnitude smaller than the size of the salt cavern in the macroscopic simulations. 

Over the past few decades, meshfree simulation methods have emerged as an alternative to the conventionally used mesh-based simulation procedures, especially in the context of modeling fluid flow. The advantages of meshfree methods are most notable for applications with complex domains, or those with moving geometry parts, free surfaces, phase boundaries, or large deformations. While modeling each of the latter cases with mesh-based methods, the highly dynamic nature of the simulations often requires an expensive global remeshing procedure. On the other hand, moving Lagrangian and semi-Lagrangian procedures fit in naturally with meshfree methods, making the simulation of dynamic geometries or phase boundaries a lot easier. In the application at hand, the modeling of the changing domain during the long-term evolution of the salt cavern falls into this category. We thus choose a meshfree approach.

In this paper, we use a meshfree Generalized Finite Difference Method (GFDM) \cite{Fan2018,Gavete2017,Katz2010,Luo2016} based on a cloud of numerical points. This method has already been successfully applied in various CFD and continuum mechanics applications. Prominent examples include water crossing of cars, water turbines, hydraulic valves, soil mechanics, metal cutting, and mold filling \cite{Jefferies2015,Kuhnert2019,Michel2017,Saucedo2019,Uhlmann2020,Uhlmann2013}. The methods presented here are part of the in-house developed software MESHFREE\footnotemark , 
which combines the advantages of GFDM and the fast linear solvers of SAMG \cite{Nick2019}. 
\footnotetext{https://www.meshfree.eu}
% Do we still want these lines:?
% has eliminated previous shortcomings concerning robust and scalable solutions of sparse, linear systems. (removed because past shortcomings need not be highlighted) 
% (Removed for repititveness) In this contribution, we present its capabilities with respect to the simulation of solution mining processes on microscopic and macroscopic scales.

We start by using a Lagrangian formulation where the discretizing point cloud moves according to the flow velocity \cite{Jefferies2015,Kuhnert2014}. This results in an accurate and natural transport of physical information. The basic physical model consists of the conservation equations for mass, momentum, and energy. For solution mining processes, we extend it by the standard $k$-$\varepsilon$ turbulence model and equations for the concentration of the occurring species (see Sect.~\ref{sec:PhysicalModel}). The GFDM specific numerics are presented in Sect.~\ref{sec:NumericalModel} with special focus on the Lagrangian and Eulerian formulations. Microscopic simulations are presented in Sect.~\ref{sec:MicroscopicScale}, and are used to determine the necessary effective model parameters of the macroscopic problem which follows. For macroscopic simulations, the Lagrangian formulation leads to a significant restriction of the time step size due to the explicit movement of the point cloud. To enable simulations in reasonable time, an Eulerian formulation should be preferred in this context. Here, the point cloud is fixed and convective terms represent the transport of physical information. The movement of the boundary of the salt cavern is based on the solution rate of the salt in the flowing water. To accurately handle this moving boundary, interior points close to the boundary are subject to an ALE-approach (Arbitrary Lagrangian-Eulerian) according to \cite{Hirt1974}. This procedure gives rise to covering the complete life cycle of a salt cavern -- several decades -- by a meshfree simulation. In Sect.~\ref{sec:MacroscopicScale}, we demonstrate the advantages of the Eulerian formulation for a simplified macroscopic example of a double-well solution mining process, followed by concluding remarks in Sect.~\ref{sec:Conclusion}.

%%%
\section{Physical model}
\label{sec:PhysicalModel}

In this section, we describe the basic physical flow model and its extensions for modeling solution mining processes, in both the macroscopic as well as microscopic simulations. Specific models needed for the density, viscosity, and heat capacity of a solution are also discussed.

\subsection{Basic equations}
\label{subsec:BasicEquations}

The basic underlying physical model is given by the conservation equations of mass, momentum, and energy in Lagrangian formulation.
\begin{align}
	\frac{d\rho}{dt} + \rho \cdot \nabla^\mathrm{T}\mathbf{v} =& 0,\label{eq:conservationequations}\\
	\frac{d}{dt}(\rho \cdot \mathbf{v}) + (\rho \cdot \mathbf{v}) \cdot \nabla^\mathrm{T}\mathbf{v} =& (\nabla^\mathrm{T} \mathbf{S})^\mathrm{T} - \nabla p + \rho \cdot \mathbf{g},\nonumber\\
	\frac{d}{dt}(\rho\cdot E) + (\rho\cdot E)\cdot\nabla^\mathrm{T}\mathbf{v} =& \nabla^\mathrm{T}(\mathbf{S}\cdot\mathbf{v})-\nabla^\mathrm{T}(p\cdot\mathbf{v}) \nonumber\\
	&+ \rho\cdot\mathbf{g}^\mathrm{T}\cdot\mathbf{v} + \nabla^\mathrm{T}(\lambda\cdot\nabla T),\nonumber
\end{align}
for  density $\rho$, velocity $\mathbf{v}\in \mathbb{R}^3$, stress tensor $\mathbf{S}\in \mathbb{R}^{3 \times 3}$ (deviatoric part, i.e.~$\mathrm{tr(\mathbf{S})=0}$), pressure $p$, body forces $\mathbf{g}\in\mathbb{R}^3$, total energy $E=c_\mathrm{v}\cdot T+\frac{1}{2}\cdot(\mathbf{v}^\mathrm{T} \cdot \mathbf{v})$, heat capacity $c_\mathrm{v}$, temperature $T$, and heat conductivity $\lambda$. Further, $\frac{d}{dt}=\frac{\partial}{\partial t} + \mathbf{v}^\mathrm{T}\nabla$ denotes the material derivative, and $\nabla=(\frac{\partial}{\partial x},\frac{\partial}{\partial y},\frac{\partial}{\partial z})^\mathrm{T}$ denotes the nabla operator.

In general, the stress tensor is split into its viscous and solid parts by $\mathbf{S} = \mathbf{S}_\mathrm{visc} + \mathbf{S}_\mathrm{solid}$ \cite{Jefferies2015,Kuhnert2014}. For the present application, the stress tensor is  purely viscous, $\mathbf{S}_\mathrm{solid} = \mathbf{0}$. The viscous part is defined by
\begin{align}
	\mathbf{S}_\mathrm{visc} = (\eta + \eta_\mathrm{turb}) \cdot \left(\nabla\mathbf{v}^\mathrm{T}+(\nabla\mathbf{v}^\mathrm{T})^\mathrm{T}-\frac{2}{3}\cdot(\nabla^\mathrm{T}\mathbf{v})\cdot\mathbf{I}\right),\label{eq:viscousstress}
\end{align}
where $\mathbf{I} \in \mathbb{R}^{3 \times 3}$ is the identity.

To incorporate turbulent effects, the standard $k$-$\varepsilon$ turbulence model \cite{Launder1974} is considered for turbulent kinetic energy $k$ and turbulent dissipation $\varepsilon$ 
\begin{align}
	\frac{dk}{dt} =& \frac{1}{\rho} \cdot \nabla^\mathrm{T}\left(\left(\eta+\frac{\eta_\mathrm{turb}}{\sigma_k}\right) \cdot \nabla k\right)-\varepsilon + \frac{1}{\rho} \cdot (P_\mathrm{pr}+P_\mathrm{b}),\label{eq:turbulencemodel}\\
	\frac{d\varepsilon}{dt} =& \frac{1}{\rho} \cdot \nabla^\mathrm{T}\left(\left(\eta+\frac{\eta_\mathrm{turb}}{\sigma_\varepsilon}\right) \cdot \nabla \varepsilon\right) -C_{2\varepsilon} \cdot \frac{\varepsilon^2}{k}\nonumber\\
	& + \frac{1}{\rho} \cdot C_{1\varepsilon} \cdot \frac{\varepsilon}{k} \cdot (P_\mathrm{pr}+C_{3\varepsilon} \cdot P_\mathrm{b}),\nonumber
\end{align}
where $\eta$ is the laminar viscosity, and $\eta_\mathrm{turb}=\rho \cdot C_\eta \cdot \frac{k^2}{\varepsilon}$ is the turbulent viscosity. Fluctuating dilatation and source terms are omitted \cite{Launder1974}. The turbulent production rate is defined by $P_\mathrm{pr} = \eta_\mathrm{turb} \cdot \Vert \nabla \mathbf{v}^\mathrm{T}\Vert^2_\mathrm{M}$ with von Mises matrix norm $\Vert\cdot\Vert_\mathrm{M}$. The turbulent buoyancy is given by $P_\mathrm{b}=-\frac{1}{\rho} \cdot \frac{\eta_\mathrm{turb}}{\mathrm{Pr}_\mathrm{turb}} \cdot \frac{\partial \rho}{\partial T} \cdot (\mathbf{g}\cdot\nabla T)$. For this model, well-established values for the constants are used 
$\sigma_k = 1.0$, 
$\sigma_\varepsilon = 1.3$, 
$C_{1\varepsilon} = 1.44$, 
$C_{2\varepsilon} = 1.92$, 
$C_{3\varepsilon} = -0.33$, 
$C_\eta = 0.09$, and 
turbulent Prandtl number $\mathrm{Pr}_\mathrm{turb} = 0.85$. 
Furthermore, a logarithmic wall function is used in the vicinity of walls.

In order to simulate solution mining processes, the basic model above is extended by convection-diffusion equations to represent the different minerals or species present in the salt mixture. For the concentration $c_i$ of species $i=1,\ldots,N$ with effective diffusion coefficient $D_{i,\mathrm{eff}}$, we have
\begin{align}
\frac{dc_i}{dt} + c_i \cdot \nabla^\mathrm{T}\mathbf{v} &= \nabla^\mathrm{T}(D_{i,\mathrm{eff}} \cdot \nabla c_i).\label{eq:convectiondiffusionconcentration}
\end{align}

In the Eulerian formulation, the material derivative is replaced by its definition, i.e.~$\frac{d}{dt}=\frac{\partial}{\partial t} + \mathbf{v}^\mathrm{T}\nabla$.

\subsection{Modeling density, viscosity, and heat capacity}
\label{subsec:ModelingDensity}

Consider the general form of the equation of state
\begin{align}
	\rho = \rho( T, c_1, \ldots, c_N ),
\end{align}
where density depends on the temperature and each of the concentrations. Based on the formulation in \cite{Laliberte2009,LaliberteCooper2004}, the density of a solution of $N$ species in water is given by
\begin{align}
	\rho_\mathrm{sol}=\left(\frac{w_{\mathrm{H}_2\mathrm{O}}}{\rho_{\mathrm{H}_2\mathrm{O}}}+\sum_{i=1}^{N}\frac{w_i}{\rho_{\mathrm{apparent},i}}\right)^{-1},\label{eq:rhosol}
\end{align}
where $w_{\mathrm{H}_2\mathrm{O}}$ and $w_i$ are the mass fraction of water and species $i$, respectively. Additionally, $w_{\mathrm{H}_2\mathrm{O}}+\sum_{i=1}^N w_i=1$ has to be satisfied. The density of water is determined by the non-linear relation
\begin{align}
	&\rho_{\mathrm{H}_2\mathrm{O}}\\
	&=\frac{ \left( 
	\left( \left( 
		\left( 	\left( A_1 \cdot T + A_2 \right) \cdot T + A_3 \right) 
		\cdot T+A_4 \right) \cdot T + A_5 \right) 
	\cdot T + A_6\right ) }{1+A_7 \cdot T},\nonumber
\end{align}
with $A_1, \ldots, A_7$ and $C_0^*, \ldots, C_4^*$ defined according to \cite{LaliberteCooper2004}. The apparent density of species $i$ is given by
\begin{align}
	&\rho_{\mathrm{apparent},i}\\
	&=\frac{(C_0^* \cdot (1-w_{\mathrm{H}_2\mathrm{O}})+C_1^*) \cdot \exp\left(0.000001 \cdot (T+C_4^*)^2\right)}{(1-w_{\mathrm{H}_2\mathrm{O}})+C_2^*+ C_3^* \cdot T}.\nonumber
\end{align}

The mass fractions $w_i$ are defined by the concentrations $c_i$ as
\begin{align}
	w_i = \frac{c_i}{\sum_{i=1}^N c_i + \rho_{\mathrm{H}_2\mathrm{O}}}.
\end{align}

The viscosity of the solution, $\eta_\mathrm{sol}(T, c_1, \ldots, c_N)$, and its heat capacity $c_\mathrm{v,sol}(T, c_1, \ldots, c_N)$ are modeled in a similar manner. For the viscosity of a solution of $N$ species in water, we use a modified version of the Arrhenius equation 
\begin{align}
	\eta_\mathrm{sol} = \left( \eta_{\mathrm{H}_2\mathrm{O}} \right)^{w_{\mathrm{H}_2\mathrm{O}}}\prod_{i=1}^N \left( \eta_i \right)^{w_i},
\end{align}
where the viscosity of water depends on the temperature as
\begin{align}
	\eta_{\mathrm{H}_2\mathrm{O}}=\frac{T+246}{(0.05594\cdot T + 5.2842)T+137.37}.
\end{align}
Furthermore, the viscosity of species $i$ is given by
\begin{align}
	\eta_i = \frac{\exp\left(\frac{V_1^*(1-w_{\mathrm{H}_2\mathrm{O}})^{V_2^*}+V_3^*}{V_4^*T+1}\right)}{V_5^*(1-w_{\mathrm{H}_2\mathrm{O}})^{V_6^*}+1}
\end{align}
with constants $V_1^*,\ldots,V_6^*$ according to \cite{Laliberte2007}.

A weighted summation of the mass fractions is used to obtain the heat capacity of a solution of $N$ species in water
\begin{align}
	c_\mathrm{v,sol} = w_{\mathrm{H}_2\mathrm{O}}c_{\mathrm{v},{\mathrm{H}_2\mathrm{O}}}+\sum_{i=1}^N w_i c_{\mathrm{v},i}.
\end{align}

Furthermore, the heat capacity of species $i$ is modeled by
\begin{align}
	c_{\mathrm{v},i}=B_1^*\exp(a)+B_5^*(1-w_{\mathrm{H}_2\mathrm{O}})^{B_6^*},
\end{align}
where $a=B_2^*T+B_3^*\exp(0.01\cdot T)+B_4^*(1-w_{\mathrm{H}_2\mathrm{O}})$ and constants $B_1^*,\ldots,B_6^*$ are according to \cite{Laliberte2009}. 

We use quadratic interpolation (extrapolation) for the definition of the heat capacity of water. Assume given temperatures $T_1,T_2,T_3$, with $T_2=T_1 + \Delta T$ and $T_3=T_2+\Delta T$ ($\Delta T > 0$). The corresponding heat capacities of water $c_{\mathrm{v},T_1}$, $c_{\mathrm{v},T_2}$, and $c_{\mathrm{v},T_3}$ are also assumed given. Then, the heat capacity of water at arbitrary temperature $T$ is determined by
\begin{align}
	c_{\mathrm{v},{\mathrm{H}_2\mathrm{O}}} =& c_{\mathrm{v},T_1} + (c_{\mathrm{v},T_2}-c_{\mathrm{v},T_1})\frac{T-T_1}{T_2-T_1}\\
&+ \frac{c_{\mathrm{v},T_3}-2c_{\mathrm{v},T_2}+c_{\mathrm{v},T_1}}{2}\frac{T-T_1}{T_2-T_1}\left(\frac{T-T_1}{T_2-T_1}-1\right).\nonumber
\end{align}
$T_1,T_2,T_3$ are chosen adaptively with $\Delta T = 5\,^\circ\mathrm{C}$, depending on the value of $T$. The range of $c_{\mathrm{v},T_k}$ values, $k=1,2,3$, are taken from \cite{Laliberte2009}, which provides the values between $0\,^\circ\mathrm{C}$ and $95\,^\circ\mathrm{C}$.

We restrict the study in this paper to sodium chloride as the species of interest. All the model constants mentioned in this section for sodium chloride are summarized in Table \ref{table:ConstantsMaterialModel}. 

\begin{table}
	\caption{Model constants for the sodium chloride solution according to \cite{Laliberte2007,Laliberte2009,LaliberteCooper2004}.}\label{table:ConstantsMaterialModel}
	\begin{center}
	\begin{tabular}{c c c c c c}
	    % original values
%		\hline
%        $i$ & $A_i$ & $C_i^*$ & $V_i^*$ & $B_i^*$ \\ \hline
%        0 & & $-3.241122 \times 10^{-3}$ & & \\ \hline
%        1 & $2.8054253 \times 10^{-10}$ & $0.063635434$ & $16.2217886$ & $-0.069356$ \\ \hline
%        2 & $1.0556302 \times 10^{-7}$ & $1.013713995$ & $1.32293087$ & $-0.0782134$ \\ \hline
%        3 & $4.6170461 \times 10^{-5}$ & $0.014595102$ & $1.48485985$ & $3.84798479$ \\ \hline
%        4 & $7.9870401 \times 10^{-3}$ & $3317.348544$ & $0.00746913$ & $-11.276211$ \\ \hline
%        5 & $16.945176$ & & $30.7802008$ & $8.73187699$ \\ \hline
%        6 & $999.83852$ & & $2.05826852$ & $1.8124593$ \\ \hline
%        7 & $0.01687985$ & & & \\ \hline
        % rounded values
        		\hline
        $i$ & $A_i$ & $C_i^*$ & $V_i^*$ & $B_i^*$ \\ \hline
        0 & & $-3.2411 \times 10^{-3}$ & & \\ \hline
        1 & $2.8054 \times 10^{-10}$ & $0.0636$ & $16.2217$ & $-0.0694$ \\ \hline
        2 & $1.0556 \times 10^{-7}$ & $1.0137$ & $1.3229$ & $-0.0782$ \\ \hline
        3 & $4.6170 \times 10^{-5}$ & $0.0146$ & $1.4849$ & $3.8480$ \\ \hline
        4 & $7.9870 \times 10^{-3}$ & $3317.3485$ & $0.0075$ & $-11.2762$ \\ \hline
        5 & $16.9452$ & & $30.7802$ & $8.7319$ \\ \hline
        6 & $999.8385$ & & $2.0583$ & $1.8125$ \\ \hline
        7 & $0.0169$ & & & \\ \hline
	\end{tabular}
    \end{center}
\end{table}

%%%
\section{Numerics based on GFDM}
\label{sec:NumericalModel}

\subsection{Point cloud preliminaries}
\label{sec:PCprelim}

In the GFDM approach, the computational domain is discretized by a cloud of numerical points. The point cloud is composed of $NP = NP(t)$ number of points, which includes points in the interior of the domain, and those at the boundary. % N is used for number of point
The initial seeding of these point clouds is done by a meshfree advancing front technique, details of which can be found in \cite{Lohner1998,SeiboldThesis}. The density of the point cloud is given by a sufficiently smooth function $h=h(\mathbf{x},t)$, the so-called interaction radius or smoothing length. Thus, $h$ prescribes the resolution of the point cloud. It is also used to define the neighborhood of each point. For a point $\mathbf{x}_j$ in the point cloud, all approximations are performed using only nearby points within a distance $h$ from it. This set of nearby points is referred to as the neighborhood or support of $\mathbf{x}_j$, and is denoted by $S_j = \{ \mathbf{x}_l \, : \, \| \mathbf{x}_l - \mathbf{x}_j \|_2 \leq h(\mathbf{x}_j)  \}$. 

To ensure a sufficient quality of the point cloud, it is ensured that no two points are closer than $r_{min} h$ apart, and that every sphere of radius $r_{max}h$ in the domain has at least one point. Thus, the inter-point distance between each point and its nearest neighbor lies in the range $(r_{min} h, r_{max} h)$. We follow conventionally used values of these parameters in Lagrangian meshfree GFDM literature, and set $r_{min} = 0.2$ and $r_{max} = 0.4$ \cite{Drumm2008,Suchde2017_CCC}. This results in about $40 - 50$ points in each interior neighborhood, with lesser at and near the boundary. 

For the Lagrangian and ALE formulations, the movement of (parts of) the point cloud with the fluid velocity can result in the minimum and maximum inter-point distance criteria being violated. This happens in form of accumulation or scattering of points which would reduce the quality of the numerical results. To prevent this, points are added in holes containing insufficient points, and are merged in regions of accumulation. This method of fixing distortion is entirely local, and much cheaper than the remeshing done in mesh-based methods. Details about these procedures of adding and deleting points follow from \cite{Drumm2008,Kuhnert2014,Kuhnert2019,SeiboldThesis,Suchde2019_MovingSurfaces}. 

%There is a direct correlation between the point cloud resolution and the quality of the simulation results (convergence of the approximation error in $h$ of order 3 for test functions/monomials up to order 2).

%%
\subsection{Differential operators}
\label{sec:DiffOps}

GFDMs generalize classical finite differences to arbitrarily spaced point clouds, using a specialized weighted moving least squares approach. Consider a function $\phi$ defined on each point of the point cloud. At each point $\mathbf{x}_j$, numerical derivatives of $\phi$ are defined as a linear combination of function values in it's neighborhood
\begin{align}
	\partial^* \phi(\mathbf{x}_j) \approx \tilde{\partial}^*_j \phi = \sum_{l \in S_j} c_{jl}^* \phi_l , \label{Eq:DiffOp}
\end{align}
where $* = x,y,z,\Delta,\dots$ denotes the derivative of interest, $\partial^*$ is the continuous differential operator, $\tilde{\partial}^*_j$ is the numerical differential operator at point $\mathbf{x}_j$, and $\phi_l = \phi(\mathbf{x}_l)$. The numerical differential operators are thus given by the coefficients $c_{jl}^*$, which are independent of the function being differentiated. They are computed by a norm minimization process that ensures that monomials up to a specified order are differentiated exactly.
\begin{align}
	&\sum_{l\in S_j}c_{jl}^{*}m_l = \partial^*_j m ,\qquad \forall m\in\mathcal{M},\\
	& \text{min }  \sum_{l\in S_j} \left( \frac{c_{jl}^{*}}{W_{jl}} \right)^2, \nonumber
\end{align}
where $\mathcal{M}$ is the set of monomials being differentiated exactly. To compute the Laplacian, the monomials are complemented by the delta function to control the central stencil value $c_{jj}^{\Delta}$, which improves stability in the pressure Poisson equations \cite{Suchde2018_Thesis}. In the present work, we consider monomials up to the order of $2$. The weighting function $W$ is defined such that neighboring points with the smallest distance to the considered point obtain the highest weight. In the present work, we use a truncated Gaussian weighting function
\begin{align}
	W_{jl} =
	\begin{cases}
			\exp \left( -c_W \frac{\| \mathbf{x}_j - \mathbf{x}_l \|^2 }{h_j^2 + h_l^2} \right), \, & \text{if } \mathbf{x}_l \in S_j \\
			0, & \text{elsewhere}
	\end{cases}
\end{align}
for a constant $c_W > 0$. We note that the same differential operators as defined above can also be equivalently derived by minimizing errors in Taylor expansions \cite{Suchde2018_Thesis}. 

Using this procedure, we compute numerical gradient operators, and a numerical Laplacian. For more details on the computation of the differential operators, we refer to \cite{Drumm2008,Kuhnert2019,Suchde2018_Thesis}.

\subsection{Time integration}
\label{subsec:TimeIntegration}

\subsubsection{Lagrangian formulation}
\label{sec:LagrangianFormulation}

A strong form discretization of the physical model (Sect.~\ref{sec:PhysicalModel}) is done using the numerical differential operators defined above, and a chosen time integration scheme. For simplicity, the following considerations are based on a first order time integration.

Starting with the Lagrangian formulation, equations \eqref{eq:conservationequations} can be rewritten as
\begin{align}
	\frac{d\rho}{dt} &= -\rho \cdot \nabla^\mathrm{T}\mathbf{v},\label{eq:conservationequationsprimitive}\\
	\frac{d\mathbf{v}}{dt} &= \frac{1}{\rho} \cdot (\nabla^\mathrm{T} \mathbf{S})^\mathrm{T} - \frac{1}{\rho} \cdot \nabla p + \mathbf{g},\nonumber\\
	(\rho \cdot c_\mathrm{v}) \cdot \frac{dT}{dt} &= \nabla^\mathrm{T}(\mathbf{S}\cdot\mathbf{v}) - (\nabla^\mathrm{T} \mathbf{S}) \cdot \mathbf{v} -p \cdot \nabla^\mathrm{T}\mathbf{v} + \nabla^\mathrm{T}(\lambda\cdot\nabla T).\nonumber
\end{align}

To improve readability, we henceforth use the shorthand $\rho=\rho_\mathrm{sol}$ and $c_\mathrm{v}=c_\mathrm{v,sol}$.

Together with equations \eqref{eq:viscousstress}--\eqref{eq:convectiondiffusionconcentration}, this is the starting point of the numerical discretization. The continuous spatial derivatives are replaced by their least squares approximated counterparts described in Sect.~\ref{sec:DiffOps}. We consider the superscript $n+1$ to denote the next time level, and $n$ for the current one, giving the time step size $\Delta t = t^{n+1} - t^n$. Below, we explain each of the steps of the discretization in the Lagrangian formulation for the microscopic scale simulations. Most of the steps are the same also for the macroscopic scale simulations, and the few differences are explained in Sect.~\ref{sec:MacroscopicScale}.

\paragraph{Step 1. Point cloud movement}~\\

The discretization procedure begins by moving the point cloud according to a second order method \cite{Suchde2018} by
\begin{align}
\mathbf{x}^{n+1} = \mathbf{x}^n + \Delta t \cdot \mathbf{v}^n + \frac{1}{2}\frac{\mathbf{v}^n - \mathbf{v}^{n-1}}{\Delta t_{0}} \cdot (\Delta t)^2\,,
\end{align}
with previous time step size $\Delta t_0 = t^n - t^{n-1}$.

\paragraph{Step 2. Temperature}~\\

A semi-implicit time integration is then carried out to compute the new temperature $T^{n+1}$ by
\begin{align}
	(\mathbf{I}_T+\mathbf{D}_T) \cdot T^{n+1} = (\rho^n \cdot c_\mathrm{v}^n) \cdot T^n + f_T,\label{eq:Tdiscretized}
\end{align}
with
\begin{align}
	\mathbf{I}_T &= \rho^n \cdot c_\mathrm{v}^n \cdot \mathbf{I},\\
\mathbf{D}_T &= -\Delta t \cdot \tilde{\nabla}^\mathrm{T} (\lambda \cdot \tilde{\nabla}),\nonumber\\
	f_T &= \Delta t \cdot ( \tilde{\nabla}^\mathrm{T}(\mathbf{S}^n \cdot \mathbf{v}^n) - (\tilde{\nabla}^\mathrm{T} \mathbf{S}^n) \cdot \mathbf{v}^n - p^n \cdot \tilde{\nabla}^\mathrm{T}\mathbf{v}^n ),\nonumber
\end{align}
where the overhead $\sim$ indicates the discrete differential operators.

To simplify notation, the index of the points has been omitted. Equation \eqref{eq:Tdiscretized} forms a sparse linear system of equations with unknowns $T^{n+1}$ at each point of the point cloud. All sparse implicit linear systems arising in this and the coming steps are solved with a BiCGSTAB solver, without the use of a preconditioner.

\paragraph{Step 3. Concentrations}~\\

A similar procedure as that done for the temperature is carried out for the concentrations. We use a semi-implicit time integration for the concentration of each species $c_i^{n+1}$, $i=1,\ldots,N$,
\begin{align}
	(\mathbf{I}_{c_i} + \mathbf{D}_{c_i}) \cdot c_i^{n+1} = c_i^n,
\end{align}
with
\begin{align}
	\mathbf{I}_{c_i} &= (\mathbf{I} + \Delta t \cdot \tilde{\nabla}^\mathrm{T} \mathbf{v}^n),\\
	\mathbf{D}_{c_i} &= -\Delta t \cdot \tilde{\nabla}^\mathrm{T}(D_{i,\mathrm{eff}} \cdot \tilde{\nabla}).\nonumber
\end{align}

\paragraph{Step 4. $\rho$, $\eta$, and $c_\mathrm{v}$}~\\

The updated density $\rho^{n+1}$, viscosity $\eta_\mathrm{sol}^{n+1}$, as well as heat capacity $c_\mathrm{v}^{n+1}$ are then determined according to the definitions in Sect.~\ref{subsec:ModelingDensity}.

Using the updated solution viscosity, a preliminary viscosity for the momentum equation is computed as $\hat{\eta}^{n+1} = \eta_\mathrm{sol}^{n+1} + \eta_\mathrm{turb}^n$.

\paragraph{Step 5. Hydrostatic pressure}~\\

The pressure is split into its hydrostatic (body forces) and dynamic parts (movement of the fluid) as
\begin{align}
	p = p_\mathrm{hyd} + p_\mathrm{dyn}.
\end{align}

First the updated hydrostatic pressure $p_\mathrm{hyd}^{n+1}$ is computed 
\begin{align}
	\tilde{\nabla}^\mathrm{T}\left( \frac{1}{\rho^{n+1}} \cdot \tilde{\nabla} p_\mathrm{hyd}^{n+1}\right) = \tilde{\nabla}^\mathrm{T} \mathbf{g}.
\end{align}

Using the updated hydrostatic pressure, a pressure guess $\hat{p}$ is computed which will be used while computing the new velocity
\begin{align}
	\hat{p} = p_\mathrm{hyd}^{n+1} + p_\mathrm{dyn}^n.
\end{align}

\paragraph{Step 6. Coupled velocity-pressure}~\\

Time integration of the first equation in \eqref{eq:conservationequationsprimitive} provides the targeted divergence of velocity $\tilde{\nabla}^\mathrm{T} \mathbf{v}^{n+1}$. To solve for $\mathbf{v}^{n+1}$ and $p^{n+1}$ in an implicit time integration scheme, we use the penalty formulation introduced in \cite{Jefferies2015,Kuhnert2014}. %It is independent of the Reynolds number. 
Using the pressure guess defined above, we obtain the following coupled velocity-pressure-system for preliminary velocity $\hat{\mathbf{v}}^{n+1}$ and correction pressure $p_\mathrm{corr}^{n+1}$:
\begin{align}
	\left( \mathbf{I} -\frac{\Delta t}{\rho^{n+1}} \cdot \tilde{\psi}_{\hat{\eta}^{n+1}}^{n+1} \right) \cdot \hat{\mathbf{v}}^{n+1} &+ \frac{\Delta t}{\rho^{n+1}} \cdot \tilde{\nabla} p_\mathrm{corr}^{n+1} \label{eq:vp1}\\
	&= \mathbf{v}^n - \frac{\Delta t}{\rho^{n+1}} \cdot \tilde{\nabla} \hat{p}+ \Delta t \cdot \mathbf{g},\nonumber\\
	\tilde{\nabla}^\mathrm{T} \left( \frac{\Delta t_\mathrm{virt}}{\rho^{n+1}} \cdot \tilde{\nabla} p_\mathrm{corr}^{n+1} \right) &= \tilde{\nabla}^\mathrm{T} \hat{\mathbf{v}}^{n+1} - \tilde{\nabla}^\mathrm{T} \mathbf{v}^{n+1}\,,\nonumber
\end{align}
with
\begin{align}
	(\tilde{\psi}_{\hat{\eta}^{n+1}}^{n+1})^\mathrm{T} =& \tilde{\nabla}^\mathrm{T} (\hat{\eta}^{n+1} \cdot \tilde{\nabla}) (\hat{\mathbf{v}}^{n+1})^\mathrm{T} \label{eq:vp2}\\
	&+ (\tilde{\nabla}\hat{\eta}^{n+1})^\mathrm{T} \cdot (\tilde{\nabla}(\hat{\mathbf{v}}^{n+1})^\mathrm{T})^\mathrm{T}\nonumber\\
	&+ \frac{\hat{\eta}^{n+1}}{3} \cdot (\tilde{\nabla} (\tilde{\nabla}^\mathrm{T} \hat{\mathbf{v}}^{n+1}))^\mathrm{T}\nonumber\\
	&- \frac{2}{3} \cdot (\tilde{\nabla}^\mathrm{T} \hat{\mathbf{v}}^{n+1}) \cdot (\tilde{\nabla} \hat{\eta}^{n+1})^\mathrm{T}\,,\nonumber
\end{align}
and $\Delta t_\mathrm{virt} = A_\mathrm{virt} \cdot \Delta t$, $0\le A_\mathrm{virt} \le 1$. If $A_\mathrm{virt} = 1$, the scheme corresponds to an implicit Chorin projection, see \cite{Chorin1968}. Theoretically, choosing $A_\mathrm{virt} = 0$ would give the exact solution. However, the linear system is ill-conditioned and can not be solved in most cases. For $0.001 \le A_\mathrm{virt} \le 0.1$, conditioning of the linear system is sufficiently good. Furthermore, the resulting preliminary velocity features a divergence which is very close to the targeted one. We note that in equations \eqref{eq:vp1} and \eqref{eq:vp2}, the stress tensor $\mathbf{S}^{n+1}$ was determined according to equation \eqref{eq:viscousstress}.

\paragraph{Step 7. Update velocity and pressure}~\\

The updates of velocity and dynamic pressure are given by
\begin{align}
	\mathbf{v}^{n+1} &= \hat{\mathbf{v}}^{n+1} - \frac{\Delta t_\mathrm{virt}}{\rho^{n+1}} \cdot \tilde{\nabla} p_\mathrm{corr}^{n+1},\\
p_\mathrm{dyn}^{n+1} &= p_\mathrm{dyn}^n + p_\mathrm{corr}^{n+1}.\nonumber
\end{align}

\paragraph{Step 8. Turbulence}~\\

For the $k$-$\varepsilon$ turbulence model, we derive a singularity formulation from equation \eqref{eq:turbulencemodel}:
\begin{align}
	\frac{d}{dt}\left(\frac{k}{\varepsilon}\right)=&(C_{2\varepsilon}-1)+C_\eta \cdot (1-C_{1\varepsilon}) \cdot \Vert\tilde{\nabla}\mathbf{v}^\mathrm{T}\Vert^2_\mathrm{M} \cdot \left(\frac{k}{\varepsilon}\right)^2\nonumber\\
	&+ \frac{C_\eta \cdot (C_{1\varepsilon} \cdot C_{3\varepsilon}-1)}{\rho \cdot \mathrm{Pr}_\mathrm{turb}} \cdot \frac{\partial\rho}{\partial T} \cdot (\mathbf{g}\cdot\tilde{\nabla}T) \cdot \left(\frac{k}{\varepsilon}\right)^2\nonumber\\
	&+ \frac{1}{\rho} \cdot \tilde\Delta_{\eta^*}\left(\frac{k}{\varepsilon}\right), \label{eq:kbyepsilonsingular}\\
	\frac{d}{dt}\left(\frac{\varepsilon}{k}\right)=&(1-C_{2\varepsilon}) \cdot \left(\frac{\varepsilon}{k}\right)^2 + C_\eta \cdot (C_{1\varepsilon}-1) \cdot \Vert\tilde{\nabla}\mathbf{v}^\mathrm{T}\Vert^2_\mathrm{M} \nonumber\\
	&+ \frac{C_\eta \cdot (1-C_{1\varepsilon} \cdot C_{3\varepsilon})}{\rho \cdot \mathrm{Pr}_\mathrm{turb}} \cdot \frac{\partial\rho}{\partial T} \cdot  (\mathbf{g}\cdot\tilde{\nabla}T)\nonumber\\
	&+ \frac{1}{\rho} \cdot \tilde\Delta_{\eta^*}\left(\frac{\varepsilon}{k}\right),\nonumber
\end{align}
where
\begin{align}
	\tilde\Delta_{\eta^*}\left(\frac{k}{\varepsilon}\right) &= \frac{\varepsilon \cdot \tilde\Delta_{\eta_k}k - k \cdot \tilde\Delta_{\eta_\varepsilon}\varepsilon}{\varepsilon^2},\\
	\tilde\Delta_{\eta^*}\left(\frac{\varepsilon}{k}\right) &= \frac{k \cdot \tilde\Delta_{\eta_\varepsilon}\varepsilon - \varepsilon \cdot \tilde\Delta_{\eta_k}k}{k^2}\nonumber
\end{align}
with
\begin{align}
	\tilde\Delta_{\eta_k}&=\tilde\nabla^\mathrm{T}\left(\left(\eta+\frac{\eta_\mathrm{turb}}{\sigma_k}\right)\cdot\tilde{\nabla}\right),\\ \tilde\Delta_{\eta_\varepsilon}&=\tilde\nabla^\mathrm{T}\left(\left(\eta+\frac{\eta_\mathrm{turb}}{\sigma_\varepsilon}\right)\cdot\tilde{\nabla}\right).\nonumber
\end{align}
If $k,\varepsilon>0$ for all $t^n \le t \le t^{n+1}$, numerical mean values can be determined from \eqref{eq:kbyepsilonsingular}:
\begin{align}
	\left.\frac{k}{\varepsilon}\right\vert_\mathrm{m}&=\frac{1}{\Delta t}\int_{t^n}^{t^{n+1}}\frac{d}{dt}\left(\frac{k}{\varepsilon}\right)\, dt,\\
	\left.\frac{\varepsilon}{k}\right\vert_\mathrm{m}&=\frac{1}{\Delta t}\int_{t^n}^{t^{n+1}}\frac{d}{dt}\left(\frac{\varepsilon}{k}\right)\, dt.\nonumber
\end{align}
We use the mean values to avoid singularities in the discretized $k$-$\varepsilon$ turbulence model.
\begin{align}
	\frac{dk}{dt}=&\frac{\tilde\Delta_{\eta_k}k}{\rho} - \left.\frac{\varepsilon}{k}\right\vert_\mathrm{m}\cdot k + C_\eta \cdot P_{\mathrm{prb},k} \cdot \left.\frac{k}{\varepsilon}\right\vert_\mathrm{m}\cdot k,\\
	\frac{d\varepsilon}{dt}=&\frac{\tilde\Delta_{\eta_\varepsilon}\varepsilon}{\rho} - C_{2\varepsilon}\cdot \left.\frac{\varepsilon}{k}\right\vert_\mathrm{m}\cdot \varepsilon + C_{1\varepsilon} \cdot C_\eta \cdot P_{\mathrm{prb},\varepsilon} \cdot \left.\frac{k}{\varepsilon}\right\vert_\mathrm{m}\cdot \varepsilon,\nonumber
\end{align}
where
\begin{align}
	P_{\mathrm{prb},k} = \Vert\tilde{\nabla}\mathbf{v}^\mathrm{T}\Vert^2_\mathrm{M} - \frac{1}{\rho \cdot \mathrm{Pr}_\mathrm{turb}}\cdot \frac{\partial\rho}{\partial T}\cdot (\mathbf{g}\cdot\tilde{\nabla}T),\\
	P_{\mathrm{prb},\varepsilon} = \Vert\tilde{\nabla}\mathbf{v}^\mathrm{T}\Vert^2_\mathrm{M} - \frac{C_{3\varepsilon}}{\rho \cdot \mathrm{Pr}_\mathrm{turb}} \cdot \frac{\partial\rho}{\partial T}\cdot (\mathbf{g}\cdot\tilde{\nabla}T).\nonumber
\end{align}

A fully implicit time integration scheme for the turbulent kinetic energy $k^{n+1}$ can now be developed as
\begin{align}
	k^{n+1} - \frac{\Delta t \cdot \tilde\Delta_{\eta_k}k^{n+1}}{\rho} &+ \Delta t\cdot \left.\frac{\varepsilon}{k}\right\vert_\mathrm{m}\cdot k^{n+1}\\
	&-\Delta t \cdot C_\eta \cdot P_{\mathrm{prb},k}^{n+1} \cdot \left.\frac{k}{\varepsilon}\right\vert_\mathrm{m}\cdot k^{n+1} = k^n.\nonumber
\end{align}

A similar procedure is used to compute the updated turbulent dissipation
\begin{align}
	\varepsilon^{n+1} - \frac{\Delta t \cdot \tilde\Delta_{\eta_\varepsilon}\varepsilon^{n+1}}{\rho} &+ \Delta t \cdot C_{2\varepsilon}\cdot  \left.\frac{\varepsilon}{k}\right\vert_\mathrm{m}\cdot \varepsilon^{n+1}\\
	&-\Delta t \cdot C_{1\varepsilon} \cdot C_\eta \cdot P_{\mathrm{prb},\varepsilon}^{n+1} \cdot \left.\frac{k}{\varepsilon}\right\vert_\mathrm{m}\cdot\varepsilon^{n+1}=\varepsilon^n\nonumber.
\end{align}

The mean values are determined analytically. This is illustrated in detail for $\left.\frac{k}{\varepsilon}\right\vert_\mathrm{m}$. Assuming that the diffusion term $\frac{1}{\rho} \cdot \tilde\Delta_{\eta^*}\left(\frac{k}{\varepsilon}\right)$ is negligible as well as defining
\begin{align}
	x = & \frac{k}{\varepsilon}, \quad a=C_{2\varepsilon}-1,\nonumber\\
	b = &\ C_\eta \cdot (C_{1\varepsilon}-1) \cdot \Vert \tilde{\nabla}\mathbf{v}^\mathrm{T}\Vert^2_\mathrm{M} \nonumber\\
	&+ \frac{C_\eta \cdot (1-C_{1\varepsilon} \cdot C_{3\varepsilon})}{\rho \cdot \mathrm{Pr}_\mathrm{turb}} \cdot \frac{\partial\rho}{\partial T} \cdot (\mathbf{g}\cdot\tilde{\nabla}T),\nonumber
\end{align}
we can rewrite equation \eqref{eq:kbyepsilonsingular} as
\begin{align}
	\frac{dx}{dt} = a - b \cdot x^2.
\end{align}
For $x_0 = \sqrt{\frac{a}{b}}$, we obtain
\begin{align}
	&x^{n+1}\\
	=& \begin{cases}x_0\cdot \mathrm{tanh}\left(\Delta t \cdot \sqrt{a \cdot b} + \mathrm{arctanh}\left(\frac{x^n}{x_0}\right)\right), &x^n<x_0\\
	x_0, &x^n=x_0\\
	x_0\cdot \mathrm{coth}\left(\Delta t \cdot \sqrt{a \cdot b} + \mathrm{arccoth}\left(\frac{x^n}{x_0}\right)\right), &x^n>x_0\end{cases}\ .\nonumber
\end{align}

Finally, the updated turbulent viscosity is determined by
\begin{align}
	\eta_\mathrm{turb}^{n+1} = \rho^{n+1} \cdot C_\eta \cdot \frac{(k^{n+1})^2}{\varepsilon^{n+1}}.
\end{align}
\subsubsection{Eulerian formulation}

In case of the Eulerian formulation, \cite{Seifarth2018} shows that a second order time integration scheme should be applied to numerically solve transport terms of the form $\mathbf{v}^\mathrm{T} \nabla$ in the GFDM context. For this purpose, the SDIRK2 method is proposed (see \cite{Alexander1977}), which features the same stability properties as an implicit Euler time integration scheme. Furthermore, an upwind discretization by means of a MUSCL reconstruction with a Superbee limiter is used.

The majority of the steps are the same as those carried out in the Lagrangian formulation. The movement step of the Lagrangian formulation is skipped here. And the coupled velocity-pressure system is modified to the following two-step procedure:
\begin{align}
	\left( \mathbf{I}_{\hat{\mathbf{v}}^{n+\alpha}} -\frac{\alpha \cdot \Delta t}{\rho^{n+\alpha}} \tilde{\psi}_{\hat{\eta}^{n+\alpha}}^{n+\alpha} \right) \cdot \hat{\mathbf{v}}^{n+\alpha} +& \frac{\alpha \cdot \Delta t}{\rho^{n+\alpha}} \cdot \tilde{\nabla} p_\mathrm{corr}^{n+\alpha}\\
	=& \mathbf{v}^n - \frac{\alpha \cdot \Delta t}{\rho^{n+\alpha}}\cdot \tilde{\nabla} \hat{p} + \alpha \cdot \Delta t \cdot \mathbf{g},\nonumber\\
	\tilde{\nabla}^\mathrm{T} \left( \frac{\Delta t_\mathrm{virt}}{\rho^{n+\alpha}} \cdot \tilde{\nabla} p_\mathrm{corr}^{n+\alpha} \right) =& \tilde{\nabla}^\mathrm{T} \hat{\mathbf{v}}^{n+\alpha} - \tilde{\nabla}^\mathrm{T} \mathbf{v}^{n+\alpha},\nonumber
\end{align}
with
\begin{align}
	\mathbf{I}_{\hat{\mathbf{v}}^{n+\alpha}} =& (\mathbf{I} + \alpha \cdot \Delta t \cdot (\widetilde{\mathbf{v}^\mathrm{T} \nabla}) \hat{\mathbf{v}}^{n+\alpha}),\\
	(\tilde{\psi}_{\hat{\eta}^{n+\alpha}}^{n+\alpha})^\mathrm{T} =& \tilde{\nabla}^\mathrm{T} (\hat{\eta}^{n+\alpha} \cdot \tilde{\nabla}) (\hat{\mathbf{v}}^{n+\alpha})^\mathrm{T}\nonumber\\
	& + (\tilde{\nabla}\hat{\eta}^{n+\alpha})^\mathrm{T} \cdot (\tilde{\nabla}(\hat{\mathbf{v}}^{n+\alpha})^\mathrm{T})^\mathrm{T}\nonumber\\
	& + \frac{\hat{\eta}^{n+\alpha}}{3} \cdot (\tilde{\nabla} (\tilde{\nabla}^\mathrm{T} \hat{\mathbf{v}}^{n+\alpha}))^\mathrm{T}\nonumber\\
	& - \frac{2}{3} \cdot (\tilde{\nabla}^\mathrm{T} \hat{\mathbf{v}}^{n+\alpha}) \cdot (\tilde{\nabla} \hat{\eta}^{n+\alpha})^\mathrm{T},\nonumber
\end{align}
and $\alpha=1-\frac{\sqrt{2}}{2}$. Density and viscosity for the intermediate step can for instance be determined by linear interpolation between time levels $n$ and $n+1$.

In the second step, the preliminary velocity is determined as
\begin{align}
	\hat{\mathbf{v}}^{n+1} &- \Delta t \cdot \alpha \cdot \mathbf{V}(\hat{\mathbf{v}}^{n+1},p_\mathrm{corr}^{n+1})\\
	&= \mathbf{v}^n + \Delta t \cdot (1-\alpha) \cdot \mathbf{V}(\hat{\mathbf{v}}^{n+\alpha},p_\mathrm{corr}^{n+\alpha}),\nonumber\\
	\tilde{\nabla}^\mathrm{T} \left( \frac{\Delta t_\mathrm{virt}}{\rho^{n+1}} \cdot \tilde{\nabla} p_\mathrm{corr}^{n+1} \right) &= \tilde{\nabla}^\mathrm{T} \hat{\mathbf{v}}^{n+1} - \tilde{\nabla}^\mathrm{T} \mathbf{v}^{n+1}\nonumber
\end{align}
with
\begin{align}
	\mathbf{V}(\hat{\mathbf{v}}^{n+1},p_\mathrm{corr}^{n+1}) =& - \frac{1}{\rho^{n+1}} \cdot (\widetilde{\mathbf{v}^\mathrm{T} \nabla}) \hat{\mathbf{v}}^{n+1} + \frac{1}{\rho^{n+1}} \cdot \tilde{\psi}_{\hat{\eta}^{n+1}}^{n+1}\\
	& - \frac{1}{\rho^{n+1}} \cdot \tilde{\nabla} \hat{p}^{n+1} - \frac{1}{\rho^{n+1}} \cdot \tilde{\nabla} p_\mathrm{corr}^{n+1} + \mathbf{g},\nonumber\\
	\mathbf{V}(\hat{\mathbf{v}}^{n+\alpha},p_\mathrm{corr}^{n+\alpha}) =& \frac{\hat{\mathbf{v}}^{n+\alpha} - \mathbf{v}^n}{\alpha \cdot \Delta t}.\nonumber
\end{align}
\subsubsection{Further details}

For more details on the Eulerian procedure, we refer to \cite{Seifarth2018}, and for similar GFDM Eulerian formulations, we refer to \cite{Suchde2019_StaticSurfaces}. 

For numerical validations of the velocity-pressure scheme used here, their implementations within a GFDM framework, and a comparison of GFDM results with other numerical methods on benchmark problems, we refer to our earlier work \cite{Drumm2008,Jefferies2015,Kuhnert2014,Michel2017,SeiboldThesis,Seifarth2018,Suchde2017_CCC}.

%%%
\section{Microscopic scale}
\label{sec:MicroscopicScale}

To study the smaller scale (both spatially and temporally) dissolution of the salt species in the water, we consider representative geometries of the salt cavern in a so-called microscopic setup. In this section, we identify effective parameters of the dissolution process. Specifically, we compute the effective diffusion coefficient and the effective transition coefficient between water and surrounding species. These will be used later, in Sect.~\ref{sec:MacroscopicScale}, in the macroscopic procedure to simulate the overall evolution of the salt cavern. The Lagrangian formulation is used here. The time integration of the underlying equations is done as presented in Sect.~\ref{sec:LagrangianFormulation}.

For the sake of brevity, we restrict the following description to sodium chloride as the species of interest. The same procedure can directly be transferred to any other species. 

\subsection{Setup}

We consider a cylinder with diameter of $5\mathrm{m}$ and height of $10\mathrm{m}$ which is initially filled with pure water, i.e.~$c_\mathrm{NaCl}(t=0) = 0$. During the simulation, the temperature is fixed to $T_0 = 20\,^\circ\mathrm{C}$.

The roof of the cylinder acts as an inexhaustible supply of sodium chloride which is modeled by applying a Dirichlet condition with saturation concentration
\begin{align}
	c_\mathrm{NaCl}^\mathrm{s} = c_\mathrm{NaCl}^\mathrm{s}(T_0) = 357\frac{\mathrm{kg}}{\mathrm{m}^3}.
\end{align}
For the hull of the cylinder, a homogeneous Neumann condition is applied. Aiming at a quasi-steady state, the bottom of the cylinder models an outflow boundary. In the interior, we solve
\begin{align}
	\frac{d c_\mathrm{NaCl}}{dt}+ c_\mathrm{NaCl} \cdot \nabla^\mathrm{T}\mathbf{v}=\nabla^\mathrm{T} (D_\mathrm{micro} \cdot \nabla c_\mathrm{NaCl}),
\end{align}
where $D_\mathrm{micro} = D_\mathrm{laminar} + D_\mathrm{turb}$. The laminar diffusion coefficient for sodium chloride is given by $D_\mathrm{laminar} = 1.611 \cdot 10^{-9} \frac{\mathrm{m}^2}{\mathrm{s}}$ (see \cite{Flury2002}). For the turbulent part, we have 
\begin{align}
	D_\mathrm{turb} = C_\eta \cdot \frac{k^2}{\varepsilon}.
\end{align}

Standard boundary conditions (Dirichlet and Neumann) are prescribed for velocity, pressure, and the turbulent quantities. The simulation runs until a quasi-steady state is reached, which will be explained below.

\subsection{Evaluation strategy}

In order to determine the effective quantities, the cylinder is split in the axial direction (z-direction) into equal sub-cylinders $SC_j$, $j=1,\ldots,J$. These are used to estimate the mass flow. The planes between the sub-cylinders are denoted by help-planes $HP_j$, $j=1,\ldots,J-1$.

We note that the moving Lagrangian nature of the simulations means that point locations are always changing in each time step, except in the trivial case when $\mathbf{v}= 0$ which does not occur here. Thus, a true steady state never occurs. Rather, simulations run till a quasi-steady state is reached, which is determined by the averaged values of the mass flow in the sub-cylinders. A quasi-steady state is said to be reached when the relative change of the mass flow in each of the sub-cylinders is within a tolerance specified (here, $10^{-4}$) for $5$ consecutive time steps.

\subsubsection{Effective diffusion coefficient}

The mass flow of sodium chloride is given by
\begin{align}
	\frac{dm}{dt}=-D_{\mathrm{NaCl},\mathrm{eff}} \cdot \frac{\partial\bar{c}_\mathrm{NaCl}}{\partial \mathbf{n}},
\end{align}
where $\bar{c}_\mathrm{NaCl}$ is the mean concentration. The mass flow and the mean concentration in sub-cylinder $SC_j$ are determined by
\begin{align}
	\frac{dm}{dt}(SC_j)&=\frac{\int_{SC_j} c_\mathrm{NaCl} \cdot v_3\, dV_{SC_j}}{\int_{SC_j} 1\, dV_{SC_j}},\\
	\bar{c}_\mathrm{NaCl}(SC_j)&=\frac{\int_{SC_j} c_\mathrm{NaCl}\, dV_{SC_j}}{\int_{SC_j} 1\, dV_{SC_j}}.\nonumber
\end{align}

Based on the mean concentration in a sub-cylinder $SC_j$, we can approximate its normal derivative with respect to the help plane $HP_j$. This yields the effective diffusion coefficients in each sub-cylinder
\begin{align}
	D_{\mathrm{NaCl},\mathrm{eff}}(SC_j|HP_j)=-\frac{\frac{dm}{dt}(SC_j)}{\left.\frac{\partial\bar{c}_\mathrm{NaCl}}{\partial \mathbf{n}}\right\rvert_{HP_j}}, \quad j=1,\ldots,J-1.
\end{align}

Once a quasi-steady state is reached, an overall effective diffusion coefficient can be determined. To accommodate the ``quasi-steady" character of the simulation, we use a time-averaged effective diffusion coefficient, over a small time interval, and over each of the sub-cylinders. This value will later be used in the macroscopic setup. 
\subsubsection{Effective transition coefficient}

The effective transition coefficient $\gamma_{\mathrm{NaCl},\mathrm{eff}}$ is derived in a manner similar to that done for the effective diffusion coefficient $D_{\mathrm{NaCl},\mathrm{eff}}$ above.
\begin{align}
	\gamma_{\mathrm{NaCl},\mathrm{eff}}(SC_j)=-\frac{\frac{dm}{dt}(SC_j)}{c_\mathrm{NaCl}^\mathrm{s}-\bar{c}_\mathrm{NaCl}(SC_j)}, \quad j=1,\ldots,J-1.
\end{align}
Once again, the time-averaged values of the effective transition coefficient in each of the sub-cylinders at the quasi-steady state gives the overall effective transition coefficient which will be used in the macroscopic simulations in Sect.~\ref{sec:MacroscopicScale}.

\subsubsection{Effective solution rate}

With the help of $\gamma_{\mathrm{NaCl},\mathrm{eff}}$, we can define the solution rate of sodium chloride for given temperature $T_0$ by
\begin{equation}
	\label{Eq:SolutionRate}
	R_\mathrm{NaCl}(T_0)=\gamma_{\mathrm{NaCl},\mathrm{eff}}(c_\mathrm{NaCl}^\mathrm{s}-c_\mathrm{NaCl}).
\end{equation}
\subsection{Numerical results}

In the simulations carried out, we choose $J=10$ to divide the cylinder domain considered into $10$ sub-cylinders of height $1\mathrm{m}$ each. We consider several levels of resolution to study the convergence of the effective parameters being determined to resolution-independent values. The coarsest resolution used is $h=0.8\mathrm{m}$ corresponding to $70400$ points in the domain. $h$ is consecutively halved till $h=0.1\mathrm{m}$ corresponding to $20892871$ points in the domain. Several resolutions in between are also considered to better illustrate the converged values of the effective parameters. We note that the number of points mentioned here are at the initial time of the simulation ($t=0$). This number of points will slightly vary in time due to the addition and deletion of points explained in Sect.~\ref{sec:PCprelim}. 

\begin{figure*}
	\centering
	\subfloat[$t = 10\mathrm{s}$]{
		\includegraphics[trim = 350 0 50 0, clip, width=0.33\textwidth]{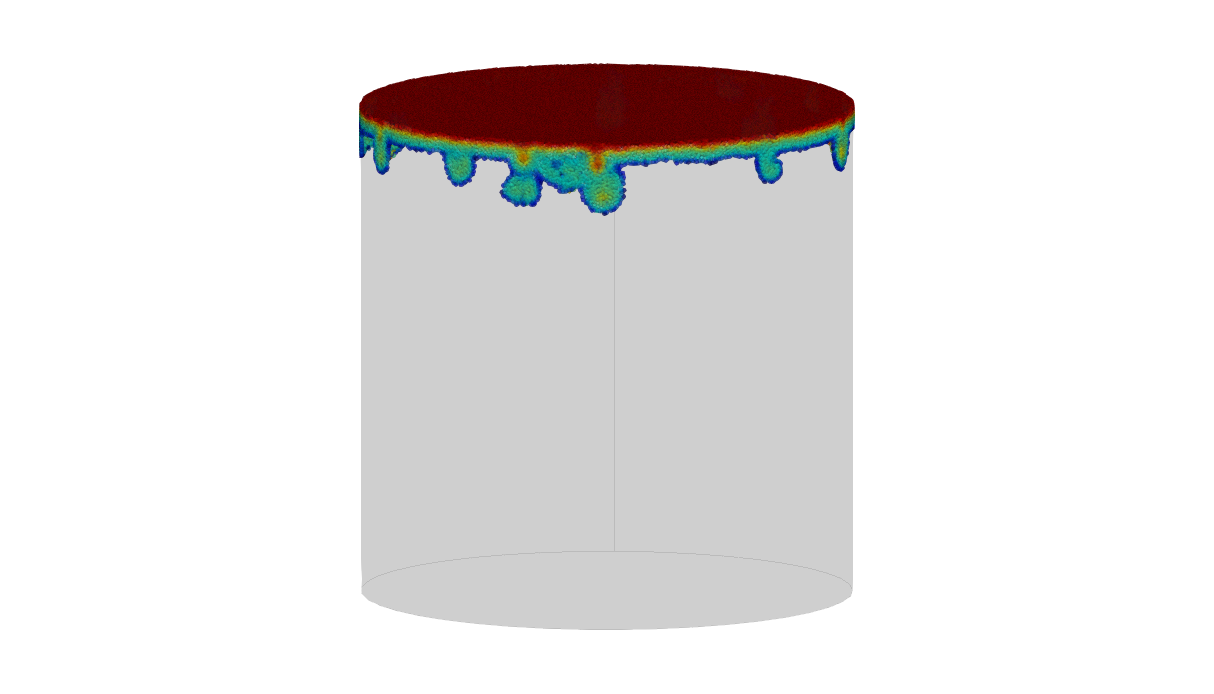}
	}
	\subfloat[$t=55\mathrm{s}$]{
		\includegraphics[trim = 350 0 50 0, clip, width=0.33\textwidth]{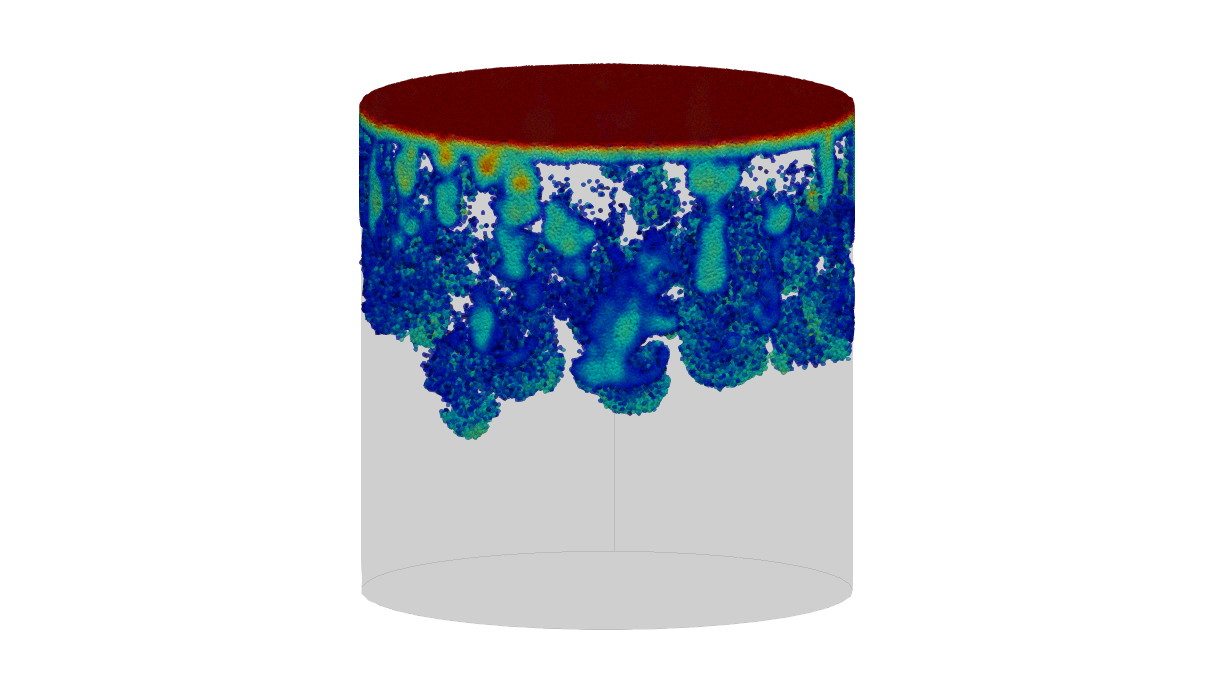}
	}
	\subfloat[$t=100\mathrm{s}$]{
		\includegraphics[trim = 350 0 50 0, clip, width=0.33\textwidth]{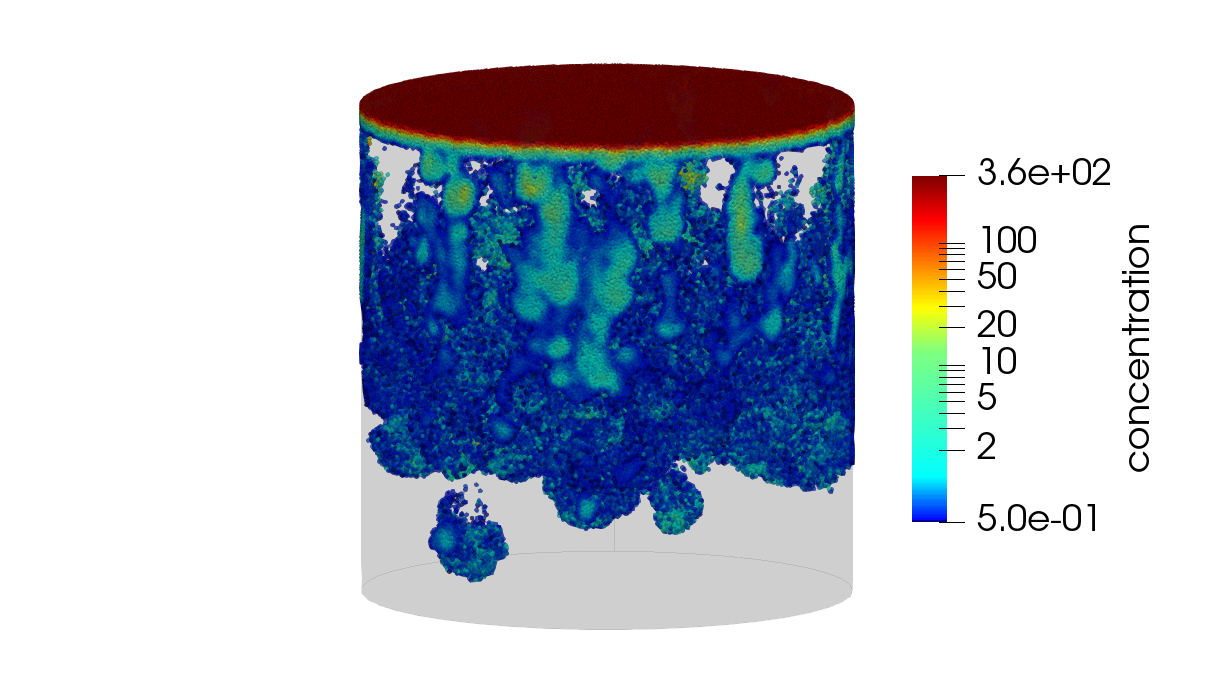}
	}
	\caption{Evolution of concentration in the microscopic simulation for interaction radius $h=0.18\mathrm{m}$ (Lagrangian formulation).}
	\label{fig:evolutionmicroscopicsimulation}
\end{figure*}

The evolution of the concentration for $h=0.18\mathrm{m}$ is illustrated in Fig.~\ref{fig:evolutionmicroscopicsimulation} in the time interval $[0\mathrm{s},100\mathrm{s}]$. As expected, the flow is characterized by viscous fingering.

\begin{figure}
	\centering
	\includegraphics[width=0.55\textwidth]{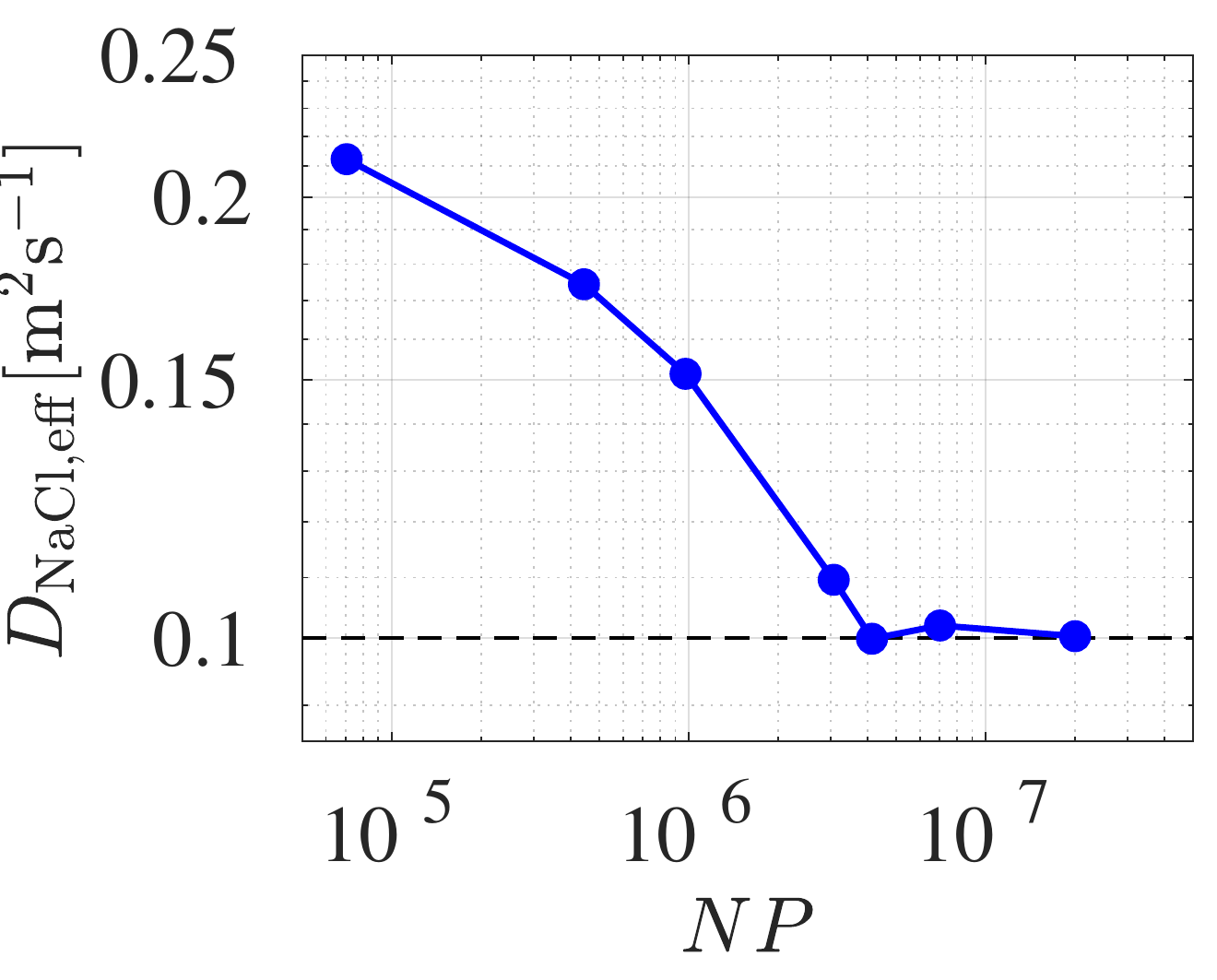}
	\caption{Convergence of the effective diffusion coefficient in the microscopic simulations. $NP$ denotes the number of the points in the initial domain.}
	\label{fig:Micro:DiffCoeff}
\end{figure}

\begin{figure}
	\centering
	\includegraphics[width=0.55\textwidth]{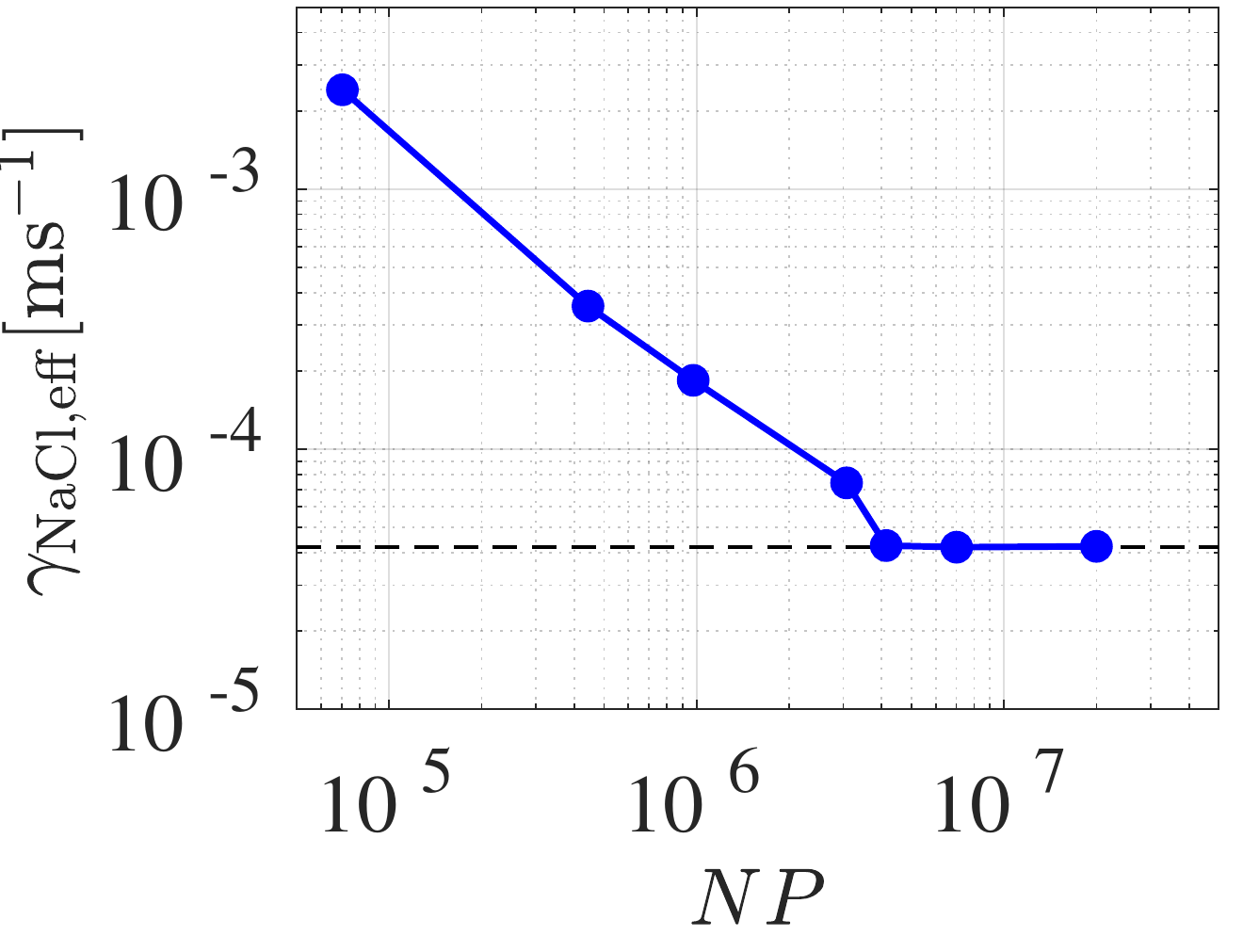}
	\caption{Convergence of the effective transition coefficient in the microscopic simulations. $NP$ denotes the number of the points in the initial domain.}
	\label{fig:Micro:TransCoeff}
\end{figure}

The convergence of effective diffusion as well as transition coefficient with decreasing $h$ is shown in Fig.~\ref{fig:Micro:DiffCoeff} and Fig.~\ref{fig:Micro:TransCoeff}, respectively. The values plotted are also tabulated in Table~\ref{table:effectivequantities}, along with the relation between the interaction radius $h$ and the number of points in the domain $NP$. The time step size is governed by $\Delta t = CFL_\mathrm{Lag} \cdot \frac{h}{|\mathbf{v}|}$, with $CFL_\mathrm{Lag}$ set to $0.2$. The diffusion coefficient converges to $D_{\mathrm{NaCl},\mathrm{eff}}=0.1\frac{\mathrm{m}^2}{\mathrm{s}}$, while the transition coefficient converges to $\gamma_{\mathrm{NaCl},\mathrm{eff}}=0.000042\frac{\mathrm{m}}{\mathrm{s}}$. Using equation \eqref{Eq:SolutionRate}, we obtain a maximum solution rate of $R_\mathrm{NaCl,max} (20\,^\circ\mathrm{C})=0.0150\frac{\mathrm{kg}}{\mathrm{m}^2\cdot\mathrm{s}}$. Compared to the solution rate of $0.0488\frac{\mathrm{kg}}{\mathrm{m}^2\cdot\mathrm{s}}$ for $T_0=23\,^\circ\mathrm{C}$ determined in \cite{Karsten1954} at a crystal level, the estimated solution rate is of the correct order of magnitude.

\begin{table}
	\caption{Estimated effective diffusion $D$ [$\frac{\mathrm{m}^2}{\mathrm{s}}$], transition coefficient $\gamma$ [$\frac{\mathrm{m}}{\mathrm{s}}$], and maximum solution rate $R$ [$\frac{\mathrm{kg}}{\mathrm{m}^2\cdot\mathrm{s}}$], with varying interaction radius $h$ [$\mathrm{m}$] and corresponding initial number of points $NP$.}\label{table:effectivequantities}
	\begin{center}
	%\begin{tabular}{*{5}{c}}
	\begin{tabular}{c r c c c}
		\hline
		$h$ & $NP$ & $D_{\mathrm{NaCl},\mathrm{eff}}$ & $\gamma_{\mathrm{NaCl},\mathrm{eff}}$ & $R_\mathrm{NaCl,max}(20\,^\circ\mathrm{C})$  \\ \hline
		$0.80$ & $70\,400$ & $0.2123$ & $2.41 \times 10^{-3}$ & $0.8604$ \\ \hline
		$0.40$ & $443\,318$ & $0.1744$ & $3.55 \times 10^{-4}$ & $0.1267$ \\ \hline
		$0.30$ & $974\,918$ & $0.1515$ & $1.84 \times 10^{-4}$ & $0.0657$ \\ \hline
%		$0.25$ & $NP$ & $0.1558$ & $1.41 \times 10^{-4}$ & $0.0504$ \\ \hline
		$0.20$ & $3\,076\,350$ & $0.1096$ & $7.42 \times 10^{-5}$ & $0.0264$ \\ \hline
%		$0.19$ & $NP$ & $0.1028$ & $5.00 \times 10^{-5}$ & $0.0179$ \\ \hline
		$0.18$ & $4\,139\,040$ & $0.0999$ & $4.26 \times 10^{-5}$ & $0.0152$ \\ \hline
		$0.15$ & $7\,014\,494$ & $0.1020$ & $4.19 \times 10^{-5}$ & $0.0150$ \\ \hline
		$0.10$ & $20\,892\,871$ & $0.1003$ & $4.23 \times 10^{-5}$ & $0.0151$ \\ \hline	
	\end{tabular}
    \end{center}
\end{table}

%%%
%%%
\section{Macroscopic scale}
\label{sec:MacroscopicScale}

We now model the overall evolution of the salt cavern during the double-well solution mining process. Both the Lagrangian as well as the Eulerian formulation are evaluated for this. 

The model equations and time integration procedures for the macroscopic scale simulations are the same as those described in Sect.~\ref{sec:PhysicalModel} and Sect.~\ref{subsec:TimeIntegration}, respectively, with a few variations. Firstly, the dissolution of the salt into the water occurs at much smaller spatial and temporal scales than those used here. To take this into account, the dissolution process of the salt at the cavern walls are modeled using a Robin boundary condition for the concentration
\begin{align}
	D_{\mathrm{NaCl},\mathrm{eff}} \cdot \frac{\partial c_\mathrm{NaCl}}{\partial \mathbf{n}}=\gamma_{\mathrm{NaCl},\mathrm{eff}} \cdot (c_\mathrm{NaCl}^\mathrm{s}-c_\mathrm{NaCl}).
\end{align}
Here, the effective diffusion coefficient $D_{\mathrm{NaCl},\mathrm{eff}}$, as well as the effective transition coefficient $\gamma_{\mathrm{NaCl},\mathrm{eff}}$ are the values determined in the microscopic simulation in Sect.~\ref{sec:MicroscopicScale}, $D_{\mathrm{NaCl},\mathrm{eff}} = 0.1\frac{\mathrm{m^2}}{\mathrm{s}}$ and $\gamma_{\mathrm{NaCl},\mathrm{eff}} = 0.000042\frac{\mathrm{m}}{\mathrm{s}}$.

A further difference in the time integration procedure comes in Steps $2$ and $4$ described in Sect.~\ref{subsec:TimeIntegration}. Here, we fix the temperature to $T_0 = 20\,^\circ\mathrm{C}$ and, subsequently, obtain the corresponding saturation concentration $c_\mathrm{NaCl}^\mathrm{s} = 357\frac{\mathrm{kg}}{\mathrm{m}^3}$. For simplicity, the following linearized relations for density and viscosity of the solution are used (see \cite{Seifarth2018})
\begin{align}
	\rho(c_\mathrm{NaCl}) &\approx (0.56 \cdot c_\mathrm{NaCl} + 1000) \frac{\mathrm{kg}}{\mathrm{m}^3},\\
	\eta(c_\mathrm{NaCl}) &\approx (1.96 \cdot 10^{-6} \cdot c_\mathrm{NaCl} + 10^{-3}) \frac{\mathrm{Pa}}{\mathrm{s}}.\nonumber
\end{align}
\subsection{Setup}

\begin{figure}
	\centering
	\includegraphics[width=0.8\textwidth]{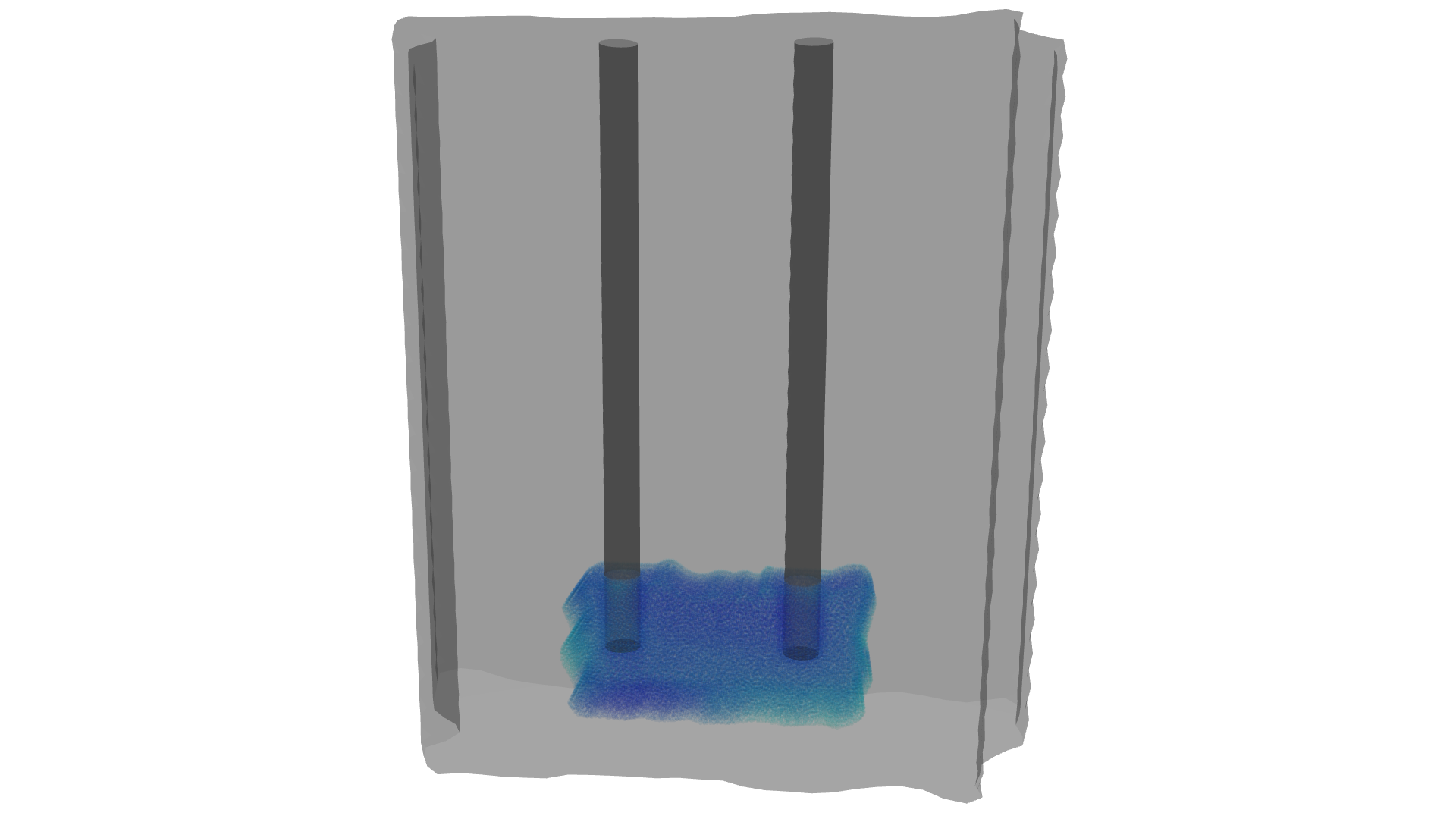}
	\caption{Macroscopic simulation setup -- initial geometry, see \cite{Seifarth2018}.}
	\label{fig:macroscopicsetup}
\end{figure}

We are interested in the geometrical evolution of the double-well salt cavern. The initial geometry is given by a small cavern filled with pure water that is surrounded by sodium chloride, see Fig.~\ref{fig:macroscopicsetup}. The dimensions of the initial cavern are approximately: width of $90\mathrm{m}$, height of $50\mathrm{m}$, and depth of $26\mathrm{m}$. The sodium chloride deposit is limited to impermeable surrounding rock. The pipe on the left side acts as an inlet of fresh water with inflow velocity $|\mathbf{v}_\mathrm{in}| = 1\frac{\mathrm{m}}{\mathrm{s}}$, whereas the pipe on the right side acts as the outlet.

In reality, the maximum diameter of the pipes is of the order of $1\mathrm{m}$. Hence, the resolution of the point cloud close to the inlet and the outlet has to be of the order of $0.1\mathrm{m}$ to ensure accurate results in case of the Lagrangian formulation. This would lead to an extremely small time step size compared to the desired simulation time of several years/decades due to the CFL-condition
\begin{align}
	\Delta t_\mathrm{Lag} \le CFL_\mathrm{Lag} \cdot \frac{h_\mathrm{min}}{|\mathbf{v}|}.
\end{align}

Numerically, we observe that stable results are achieved for $CFL_\mathrm{Lag}=0.15$, which results in $\Delta t_\mathrm{Lag} = \mathcal{O}(0.1\mathrm{s})$. %Since $|\mathbf{v}|\ge |\mathbf{v}_\mathrm{in}|$ and $h_\mathrm{min} = \mathcal{O}(0.1\mathrm{m})$, we obtain $\Delta t_\mathrm{Lag} = \mathcal{O}(0.1\mathrm{s})$.
Performing long-term simulations over months of simulation time is not feasible with such a small time step. This results in the need for using the Eulerian formulation for the problem at hand. 

In order to allow for a comparison of Lagrangian and Eulerian formulation, we consider pipes of diameter $12\mathrm{m}$. This also decreases the required actual time being simulated. Numerically, we observe that with the significantly larger diameter used here, the evolution of the cavern only requires a few hours of physical time to be simulated, compared to the few months or years with the actual diameter. We note that despite this time reduction, this still corresponds to a time scale two orders of magnitude greater than that used in the microscopic simulations in Sect.~\ref{sec:MicroscopicScale}.

The simulations are performed with a constant interaction radius of $h=4\mathrm{m}$. An important point to note here is that this spatial resolution considered is of the same order of magnitude as the height of the sub-cylinders in the microscopic simulations in Sect.~\ref{sec:MicroscopicScale}. 

\subsection{Movement of the boundary}

The movement of the boundary of the cavern can be defined by the Stefan condition
\begin{align}
	\rho v^\star = \gamma_{\mathrm{NaCl},\mathrm{eff}}(c_\mathrm{NaCl}^\mathrm{s}-c_\mathrm{NaCl}),
\end{align}
see \cite{Javierre2003}. This yields
\begin{align}
	v^\star = \frac{\gamma_{\mathrm{NaCl},\mathrm{eff}}}{\rho} \cdot (c_\mathrm{NaCl}^\mathrm{s}-c_\mathrm{NaCl})
\end{align}
and, consequently, a movement of the boundary in normal direction $\mathbf{n}$ with velocity $\mathbf{v}_\mathrm{boundary} = v^\star \cdot \mathbf{n}$. To speed up computation, a time lapse procedure can be applied \cite{Seifarth2018}. Due to small flow velocities inside the salt cavern, an additional speed-up factor $A$ can be introduced in the definition of $v^\star$ by
\begin{align}
    v^\star_A = \frac{A \cdot \gamma_{\mathrm{NaCl},\mathrm{eff}}}{\rho} \cdot (c_\mathrm{NaCl}^\mathrm{s}-c_\mathrm{NaCl}).
\end{align}

For stability reasons, we require
\begin{align}
\Delta t \cdot v^\star_A \le 0.8\cdot h.\label{eq:stabilitycriterionDeltatvStarA}
\end{align}

The maximum movement velocity of the boundary occurs in case of pure water, i.e.~$c_\mathrm{NaCl}=0$, and is given by
\begin{align}
v^\star_{A,\mathrm{max}} = \frac{A\cdot \gamma_{\mathrm{NaCl},\mathrm{eff}} \cdot c_\mathrm{NaCl}^\mathrm{s}}{\rho}.
\end{align}

Given a maximum time step size $\Delta t_\mathrm{max}$, equation \eqref{eq:stabilitycriterionDeltatvStarA} leads to the constraint
\begin{align}
A \le \frac{0.8\cdot h \cdot \rho}{\Delta t_\mathrm{max}\cdot \gamma_{\mathrm{NaCl},\mathrm{eff}} \cdot c_\mathrm{NaCl}^\mathrm{s}}.
\end{align}

\begin{figure}
	\centering
	\includegraphics[width=0.8\textwidth]{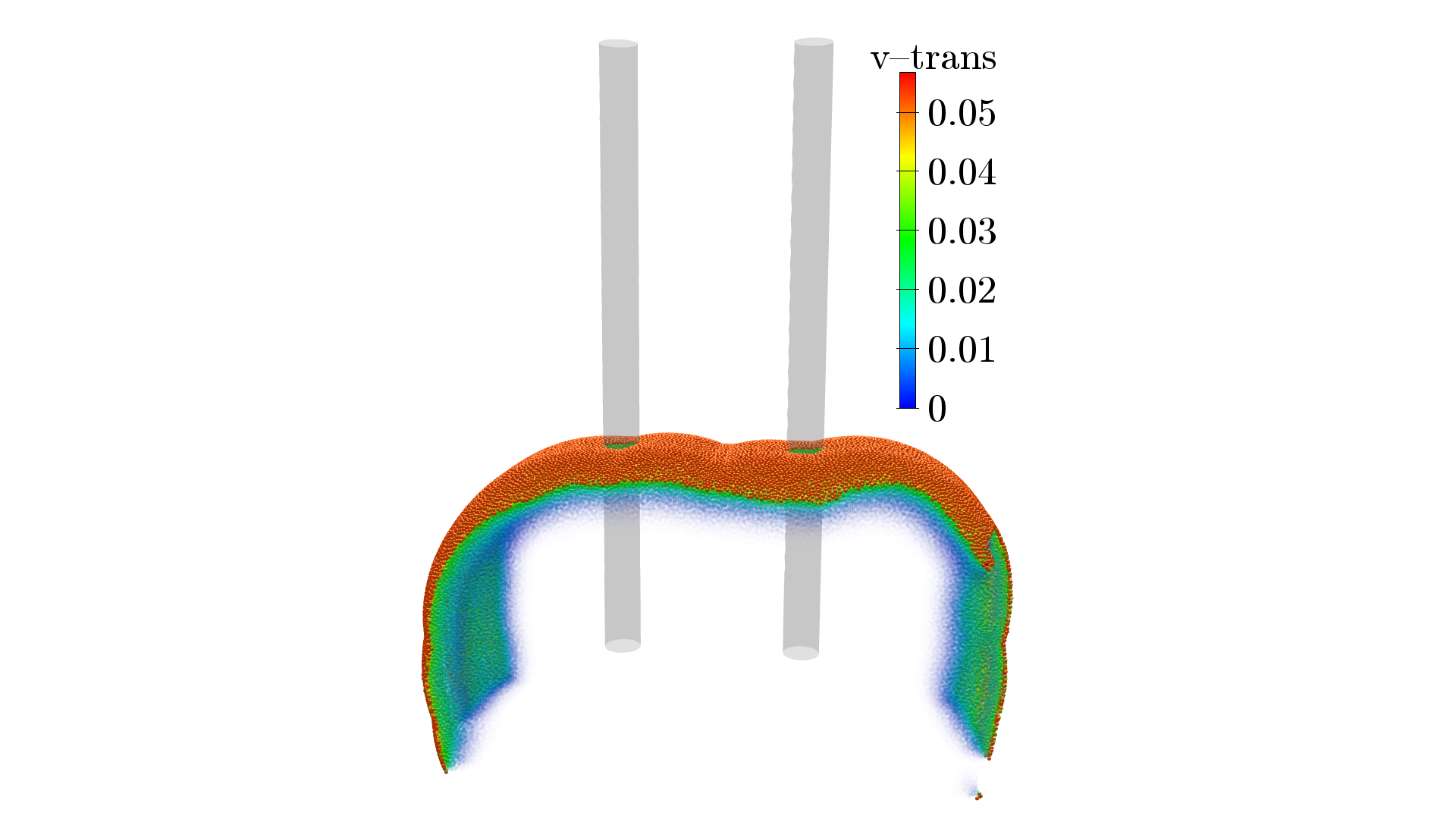}
	\caption{Translational velocity in the macroscopic simulation, see \cite{Seifarth2018} (Eulerian formulation including ALE at the moving boundary).}
	\label{fig:TranslationalVelocity}
\end{figure}

Due to the movement of the boundary, interior points close to this boundary have to move in the Eulerian formulation also. For this purpose, the ALE-approach presented in \cite{Hirt1974} is used. Based on current and future position of an affected interior point, the translational velocity
\begin{align}
	\mathbf{v}_\mathrm{trans} = \frac{\mathbf{x}^{n+1}-\mathbf{x}^n}{\Delta t}
\end{align}
is determined, see Fig.~\ref{fig:TranslationalVelocity}. Due to the explicit movement of these points, the convection terms in the numerical model in Eulerian form in Sect.~\ref{subsec:TimeIntegration} must refer to the relative velocity $\mathbf{v}-\mathbf{v}_\mathrm{trans}$ instead of $\mathbf{v}$.

Furthermore, this introduces a CFL-condition of the form
\begin{align}
\Delta t_\mathrm{ALE} \le CFL_\mathrm{ALE} \cdot \frac{h_\mathrm{min}}{v^\star}.
\end{align}

This depends on the boundary velocity $v^\star = \mathcal{O}(0.01\frac{\mathrm{m}}{\mathrm{s}})$ which is considerably smaller than the flow velocity (the inflow velocity is $1\frac{\mathrm{m}}{\mathrm{s}}$, while the maximum velocity in the domain is even bigger). $h_\mathrm{min}$ is subject to the desired resolution at the moving boundary. At the inlet and the outlet, a coarse resolution is sufficient in this case.

\subsection{Numerical results}

\begin{figure*}
	\centering
	\subfloat[$t = 1537\mathrm{s}$]{
		\includegraphics[trim = 400 0 400 0, clip, width=0.3\textwidth]{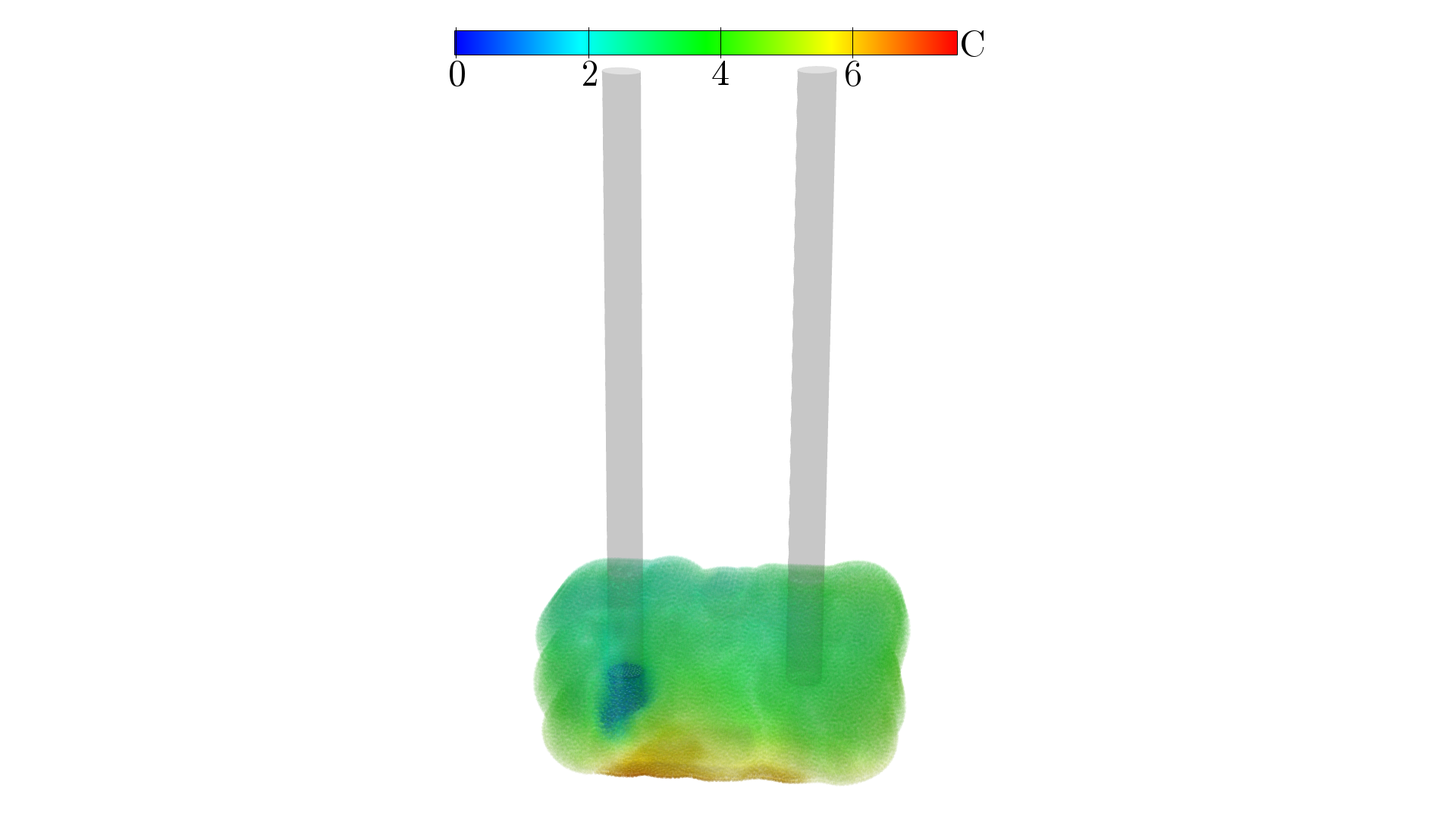}
	}
	\subfloat[$t=2446\mathrm{s}$]{
		\includegraphics[trim = 400 0 400 0, clip, width=0.3\textwidth]{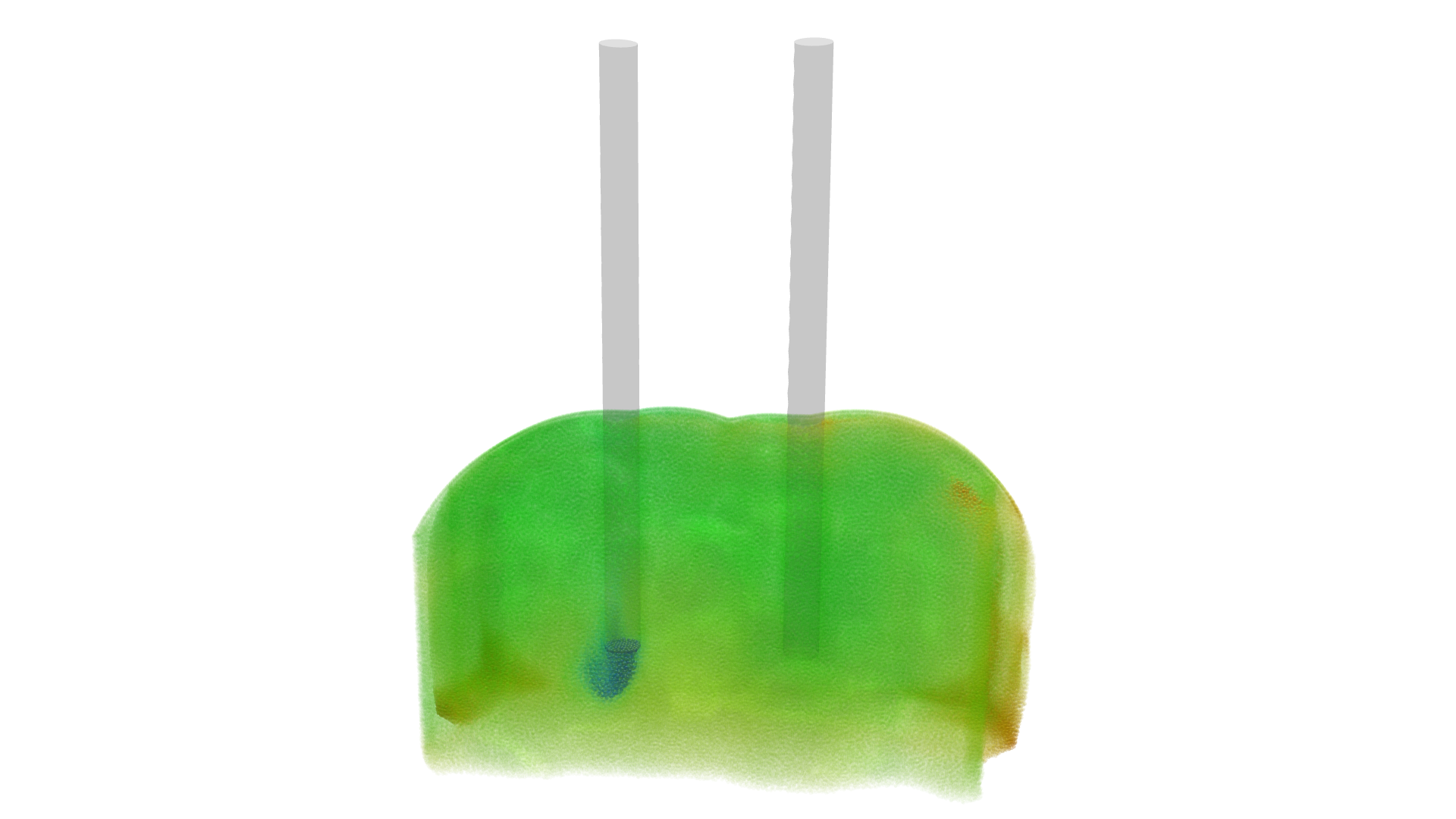}
	}
	\subfloat[$t=3228\mathrm{s}$]{
		\includegraphics[trim = 400 0 400 0, clip, width=0.3\textwidth]{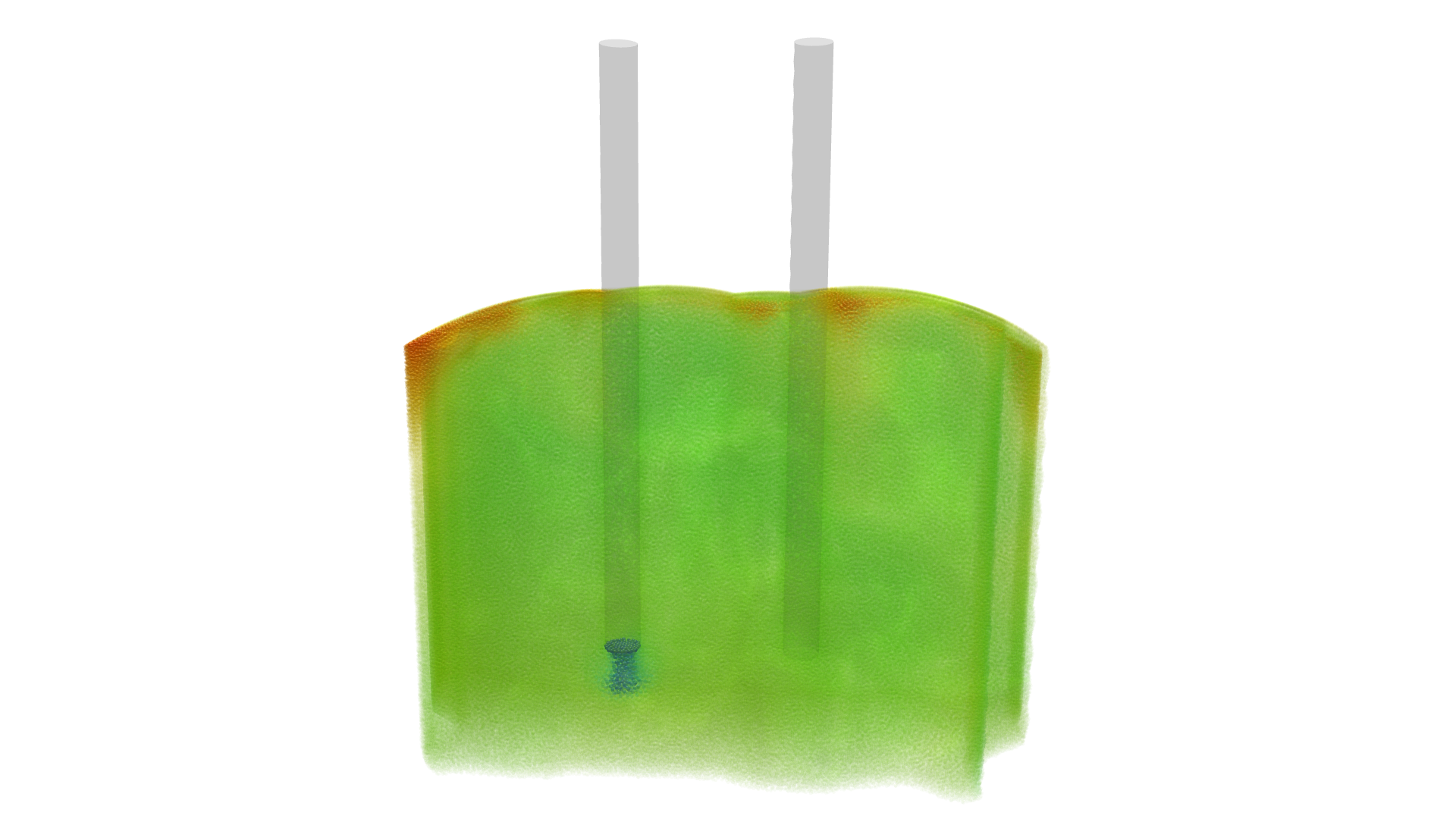}
	}
	\caption{Evolution of the macroscopic simulation for interaction radius $h=4\mathrm{m}$ -- concentration, see \cite{Seifarth2018} (Eulerian formulation including ALE at the moving boundary). }
	\label{fig:evolutionmacroscopicsimulation}
\end{figure*}

\begin{figure}
	\centering
	\includegraphics[width=0.5\textwidth]{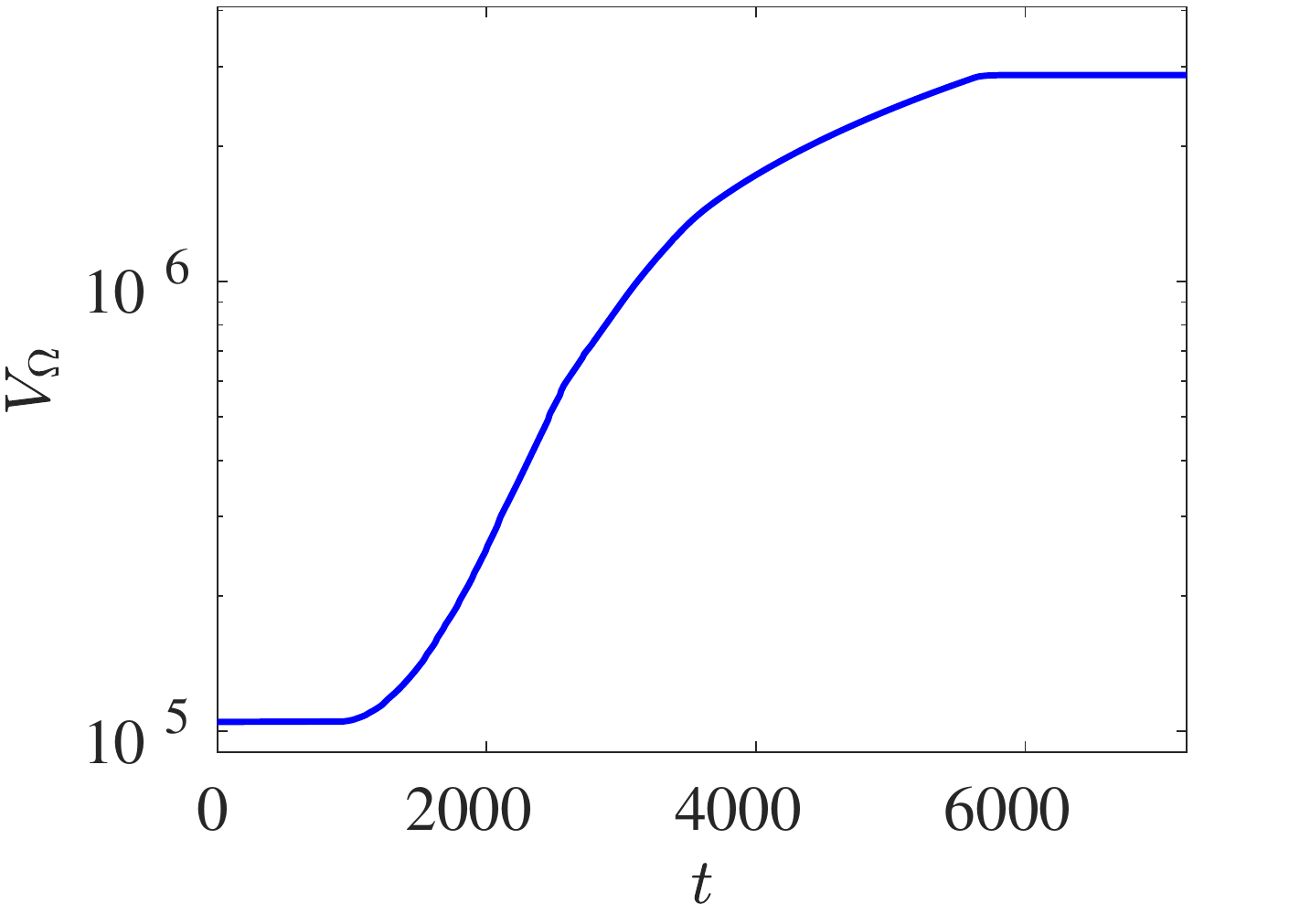}
	\caption{Expansion in the volume of the computational domain of the macroscopic simulation as the simulation progresses.}
	\label{fig:MacroVolume}
\end{figure}

Starting from the initial domain as shown in Fig.~\ref{fig:macroscopicsetup}, the salt cavern expands till the outer domain is filled. Physically, this outer domain can represent either a rock formation where the salt cavern ends, or prescribed limits of the region where the solution mining is to be carried out. Fig.~\ref{fig:evolutionmacroscopicsimulation} illustrates the evolution of the salt cavern in the Eulerian formulation according to $\mathrm{C}=c_\mathrm{NaCl}$. An animation of this process can be found in the Online Resource. This expansion is quantified by plotting the volume as a function of time in Fig.~\ref{fig:MacroVolume}. We note that once the entire outer domain is filled, the volume of the domain about $27.4$ times that of the initial domain. This shows the need of a meshfree method for the present application. If a mesh-based method were to be used for this simulation, the entire domain would need to be meshed initially, and an expensive and less accurate tracking of the expansion would need to be carried out.  

\begin{figure}
	\centering
	\includegraphics[width=0.5\textwidth]{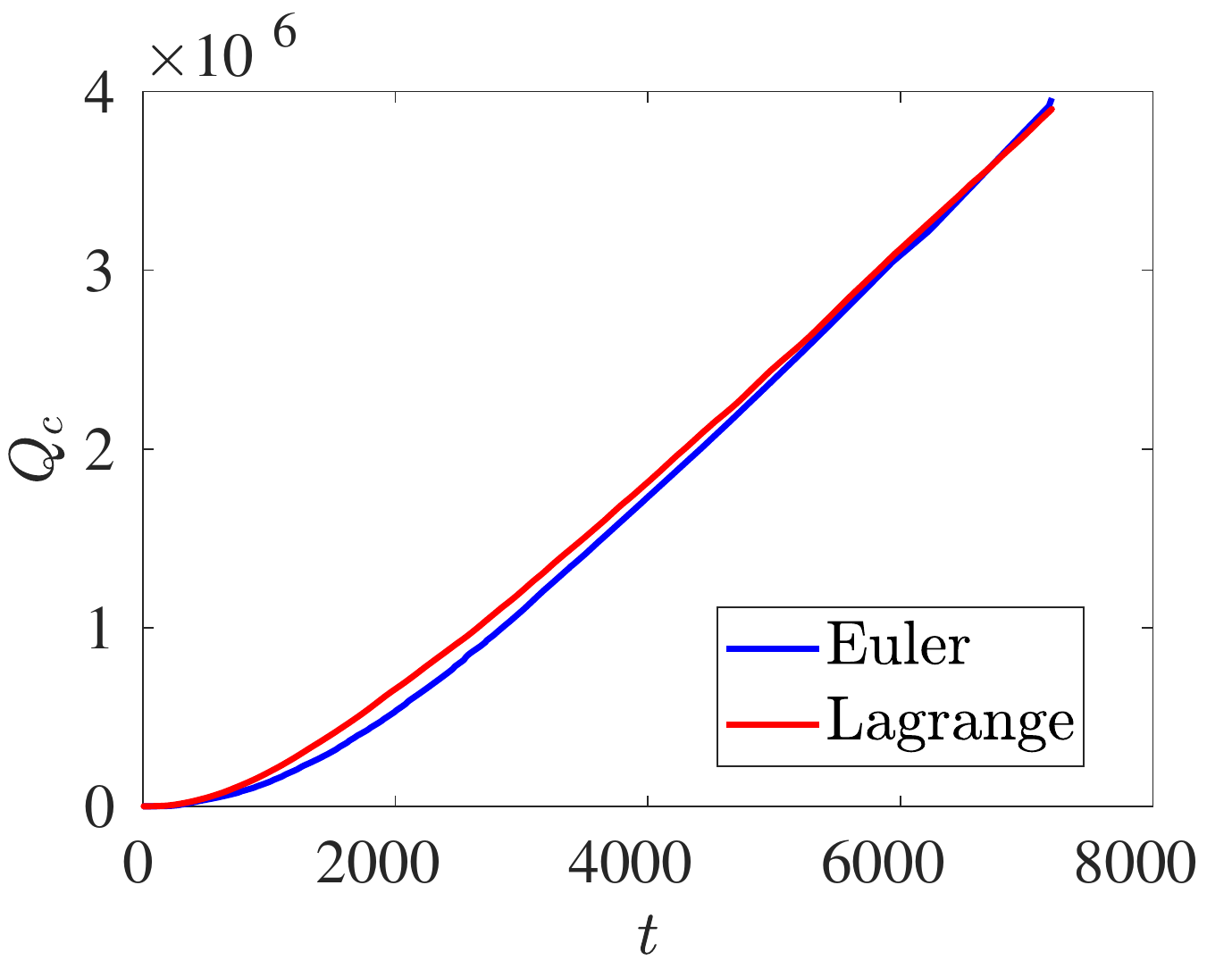}
	\caption{Comparison of the concentration flux across the outflow pipe between the Eulerian (with moving boundaries) and Lagrangian simulations.}
	\label{fig:MacroPlots}
\end{figure}

To compare the results of the Eulerian and Lagrangian formulations, we consider simulations on the same initial point cloud with $h=4\mathrm{m}$ which corresponds to $NP_0 = 90\,744$ points at the initial state. The simulations are run until $t=7200\mathrm{s} = 2 \mathrm{hours}$. At the end time, both simulations have about $1.5$ million points in the final expanded domain. To quantify the results, and to enable a comparison between the two formulations, we consider the time integration of the concentration weighted flux at the outflow boundary
\begin{equation}
	Q_c(t) = \int_0^t \left( \int_{\partial \Omega_{\mathrm{out}}} c_\mathrm{NaCl}\, \mathbf{v} \cdot \mathbf{n} \;\mathrm{d}A \right) \;\mathrm{d}\tau \,,
\end{equation}
where $\partial \Omega_{\mathrm{out}}$ is the outflow boundary located at the top of the extraction well. Physically, this represents a measure of the concentration of salt being extracted. The time evolution of $Q_c$ is shown in Fig.~\ref{fig:MacroPlots}. It illustrates that both formulations produce very similar results.

To emphasize the need of the ALE formulation for such a simulation, we compare the time steps required in both the ALE and Lagrangian formulations for stability. Considering the simulation time of $7200\mathrm{s}$, we observe that the Lagrangian formulation required at least 22915 time steps to obtain stable results, which corresponds to an average time step size of $\Delta t \approx 0.31 \mathrm{s} $. On the other hand, similar results, as shown in Fig.~\ref{fig:MacroPlots}, can be obtained in the Eulerian formulation (with ALE near the boundaries) with only 936 time steps corresponding to an average time step size of $\Delta t \approx 7.69 \mathrm{s}$, which is approximately $25$ times that needed in the Lagrange case.

%

%%%
\section{Conclusions}
\label{sec:Conclusion}

In this contribution, we presented the capabilities of the Generalized Finite Difference Method (GFDM) implemented in the simulation software MESHFREE regarding solution mining processes on a macroscopic, as well as a microscopic scale. Both Lagrangian and Eulerian approaches were considered.

On the macroscopic scale, we considered the expansion of the salt cavern as a result of erosion occurring as the salt dissolves in the water. In reality, this procedure occurs over the time span of several months or years. A simplified geometry was considered here, which enabled a comparison between the Eulerian and Lagrangian formulations. In this simplified macroscopic set-up, the expansion of the salt cavern occurred over the time scale of several hours. Since the dissolution of salt in water occurs on a much smaller time level we also considered a microscopic set-up over a duration of a few minutes. This was used to determine effective parameters governing the dissolution process. Using the example of sodium chloride as the species of interest, effective diffusion and transition coefficients were determined in the microscopic simulations. These values were then used in the macroscopic simulations to determine the evolution of the concentration inside the salt cavern and to specify the solution rate of the salt species at the boundary, i.e.~to model the geometrical evolution of the salt cavern.

A comparison of the numerical results of the Lagrangian and Eulerian formulations (extended by an ALE-approach) in the macroscopic case illustrates the advantages of the latter one due the possibility of using much larger time step sizes. Aiming at a simulation time of several years, the forecast computation time for a simulation of a double-well solution mining process based on the Lagrangian formulation would be of the order of years. In contrast to that, the flexibility of the Eulerian formulation regarding the resolution of the point cloud (local refinement only at the moving boundary) enables meshfree simulations in reasonable time -- especially in terms of real applications.

%\begin{acknowledgements}
%If you'd like to thank anyone, place your comments here
%and remove the percent signs.
%\end{acknowledgements}

% Authors must disclose all relationships or interests that 
% could have direct or potential influence or impart bias on 
% the work: 
%
\section*{Conflict of interest}

On behalf of all authors, the corresponding author states that there is no conflict of interest.

% BibTeX users please use one of
%\bibliographystyle{spbasic}      % basic style, author-year citations
%\bibliographystyle{spmpsci}      % mathematics and physical sciences
%\bibliographystyle{spphys}       % APS-like style for physics
%\bibliography{}   % name your BibTeX data base

\bibliographystyle{abbrv}
%\bibliography{./Ref/SM}

\end{document}